\documentclass{jfm}
\usepackage{graphicx}
\usepackage{epstopdf, epsfig}
\graphicspath{ {./Figures/} }
\usepackage{color}
\definecolor{gray}{rgb}{0.5,0.5,0.5}
\usepackage{psfrag}
\usepackage{amsmath}
\usepackage{mathabx}
\usepackage{tikz}
\usetikzlibrary{shapes}
\usepackage{tabularx}
\usepackage{mathalfa}
\usepackage{multirow}
\def\spacce#1{\hskip #1pt}
\def\drawline#1#2{\raise 2.5pt\vbox{\hrule width #1pt height #2pt}}
\def\solid{\drawline{24}{.5}\nobreak}
\def\linearrow{{\Large $\rightarrow$}\raisebox{1.2pt}{\hspace{-13pt}\drawline{20}{.2}}\nobreak}

\def\bdash{\hbox{\drawline{5.8}{.5}\spacce{2}}}

\def\dashed{\bdash\bdash\bdash\nobreak}
\def\dotdashed{\bdash\spacce{3}\bdot\spacce{3}\bdash}

\def\bdot{\hbox{\drawline{1}{.5}\spacce{2}}}

\def\dotted{\hbox{\leaders\bdot\hskip 24pt}\nobreak}

\def\trian{\raise 1.25pt\hbox{$\scriptstyle\triangle$}\nobreak}

\def\dtrian{\raise 1.25pt\hbox%
{$\scriptscriptstyle\bigtriangledown$}\nobreak}

\def\squar{\raise 1.25pt\hbox{$\scriptstyle\Box$}\nobreak}

\def\diamon{\raise 1.25pt\hbox{$\scriptstyle\diamond$}\nobreak}

\def\linedtri1{\hbox{\bdash\hspace{-1.6mm}$\bigtriangleup$\hspace{-0.8mm}\bdash}\nobreak}
\def\soliddtrian1{$\blacktriangledown$\nobreak}
\def\solidrtrian2{$\blacktriangleright$\nobreak}
\def\solidltrian3{$\blacktriangleleft$\nobreak}
\def\dd{{\, \rm{d}}}

\def\beq{\begin{equation}}
\def\eeq{\end{equation}}

\def\ii{{\rm i}}

\def\aaa{{\it a}}
\def\bbb{{\it b}}
\def\ccc{{\it c}}
\def\ddd{{\it d}}

\def\ggg{{\it g}}

\def\r#1{(\ref{#1})}

\def\refb#1{{\color{black}#1}}
\def\refc#1{{\color{black}#1}}
\def\refm#1{{\color{black}#1}}

\newcommand{\linescircle}[1]{\raisebox{0.5pt}{\tikz{\draw[-,#1!40!#1,solid,line width = 1pt](0,0) -- (8mm,0);\node[draw,scale=0.5,circle,#1!40!#1,fill=#1!20!#1,line width = 0.9pt]() at (4mm,0.0mm) {};}}}

\newcommand{\linesquare}[1]{\raisebox{0.5pt}{\tikz{\draw[-,#1!40!#1,solid,line width = 1pt](0,0) -- (8mm,0);\node[draw,scale=0.5,regular polygon, regular polygon sides=4,#1!40!#1,fill=white!20!white,line width = 0.9pt]() at (4mm,0.0mm) {};}}}

\newcommand{\linestriau}[1]{\raisebox{0.5pt}{\tikz{\draw[scale=1,-,#1!40!#1,solid,line width = 1pt](0,0.7pt) -- (8mm,0.7pt);\node[draw,scale=0.35,regular polygon, regular polygon sides=3,#1!40!#1,fill=#1!20!#1,line width = 0.9pt]() at (4mm,0.0mm) {};}}}

\newcommand{\linestriar}[1]{\raisebox{0.5pt}{\tikz{\draw[scale=1,-,#1!40!#1,solid,line width = 1pt](0,-0.7pt) -- (8mm,-0.7pt);\node[draw,scale=0.35,regular polygon, regular polygon sides=3,rotate=-90,#1!40!#1,fill=#1!20!#1,line width = 0.9pt]() at (4mm,0.0mm) {};}}}

\newcommand{\dashtriau}[1]{\raisebox{0.5pt}{\tikz{\draw[scale=1,-,#1!40!#1,dashed,line width = 1pt](0,-0.7pt) -- (8mm,-0.7pt);\node[draw,scale=0.35,regular polygon, regular polygon sides=3,#1!40!#1,fill=#1!20!#1,line width = 0.9pt]() at (4mm,0.0mm) {};}}}

\newcommand{\dashtriar}[1]{\raisebox{0.5pt}{\tikz{\draw[scale=1,-,#1!40!#1,dashed,line width = 1pt](0,-0.7pt) -- (8mm,-0.7pt);\node[draw,scale=0.35,regular polygon, regular polygon sides=3,#1!40!#1,rotate=-90,fill=#1!20!#1,line width = 0.9pt]() at (4mm,0.0mm) {};}}}

\newcommand{\dashquare}[1]{\raisebox{0.5pt}{\tikz{\draw[-,#1!40!#1,dashed,line width = 1pt](0,0) -- (8mm,0);\node[draw,scale=0.5,regular polygon, regular polygon sides=4,#1!40!#1,fill=#1!20!#1,line width = 0.9pt]() at (4mm,0.0mm) {};}}}

\DeclareTextFontCommand{\textarial}{}
\definecolor{orange}{RGB}{255,127,0}

\shorttitle{Entropy, irreversibility and cascades}
\shortauthor{Alberto Vela-Mart\'in and Javier Jim\'enez}

\title{Entropy, irreversibility and cascades in the inertial range of isotropic turbulence}

\author{Alberto Vela-Mart\'in%\aff{1}
  \corresp{\email{albertovelam@gmail.com}},
  Javier Jim\'enez}%\aff{1}}

\affiliation{School of Aeronautics, Universidad Polit\'ecnica de Madrid, 28040 Madrid, Spain}

\begin{document}

\maketitle

\begin{abstract}
This paper analyses the turbulent energy cascade from the perspective of statistical
mechanics, and relates inter-scale energy fluxes to \refm{statistical irreversibility} and information-entropy production.
The microscopical reversibility of the energy cascade is tested by constructing a reversible
3D turbulent system using a dynamic model for the sub-grid stresses. This system, when reversed
in time, develops a sustained inverse cascade towards the large scales, evidencing that the
characterisation of the inertial energy cascade must consider the possibility of an inverse regime.
This experiment is used to study the origin of statistical irreversibility and the prevalence of direct over inverse energy cascades in isotropic turbulence.
Statistical irreversibility, a property of statistical ensembles in phase space related to entropy production, is connected to the dynamics of the energy cascade in physical space by considering the space locality of the energy fluxes and their relation to the local structure of the flow.
A mechanism to explain the probabilistic prevalence of direct energy transfer is proposed based 
the dynamics of the rate-of-strain tensor, which is identified as the most important source of statistical 
irreversibility in the energy cascade.
\end{abstract}

% ------------------------------------------------------------------------------------------
\section{Introduction}

%Introduction to the problem of the cascade. How it is not clear what we refer as the cascade.

The turbulence cascade is a scientific paradigm that dates back to the beginning of the
20th century. It has been addressed in different ways, and the objects used to describe it
are many. From the classical view of \citet{richardson1922weather} to the recent
analysis of coherent structures \citep{cardesa2017turbulent}, through the
statistical theory of \citet{kolmogorov1941local} or the spectral representation
\citep{waleffe1992nature}, all descriptions focus on the same phenomena through diverse
approaches. In all cases, the cascade tries to account for the gap between the scales where
energy is produced and those at which it is dissipated. A concept of energy flux or energy
transport from large to small scales is always present. We know that this cascade must take
place because this gap must be bridged, but the mechanisms that underlie it are to date
poorly or ambiguously identified. A thorough description of these mechanisms and their
causes is needed if a complete theory of turbulence is to be constructed, and necessarily
involves exploring multiple representations of the turbulence cascade.

In this paper we expand the investigation of the energy cascade
from the perspective of dynamical system theory and statistical mechanics. 
We explore microscopical reversibility in the inertial range dynamics, 
and its implications for the representation of the turbulence cascade as a process driven by entropy production
\refc{\citep{kraichnan1967inertial}}, emphasising that the prevalence of a direct energy transfer
is probabilistic, rather than an unavoidable consequence of the equations of motion.

Turbulence is a highly dissipative and essentially irreversible phenomenon. An intrinsic
source of the irreversibility of the Navier--Stokes equations is viscosity,
which accounts for the molecular diffusion of momentum. However, when the Reynolds number is
high, only the smallest scales with locally low Reynolds number are affected. At scales much
larger than the Kolmogorov length, viscous effects are negligible and dynamics are only driven by
inertial forces, which, as we will demonstrate, generate time-reversible dynamics. This
is clear from the truncated Euler equations, which are known to be reversible and to display
features specific to fully developed turbulent flows under certain conditions
\citep{she1993constrained, cichowlas2005effective}. 
Although these equations are invariant
to a reversal of the time axis, statistical irreversibility appears as a tendency towards
preferred evolutions when the flow is driven out of equilibrium. The energy cascade is
an important manifestation of this irreversibility. Thus, we should distinguish between
intrinsic irreversibility, which is directly imposed by the equations of motion,
and statistical irreversibility, which is a consequence of the dynamical
complexity of highly chaotic systems with many degrees of freedom. In the absence of an
intrinsic irreversible mechanism, a dynamical system can be microscopically reversible and
statistically, or macroscopically, irreversible. This idea, which is widely understood in the study of dynamical
systems, has not been sufficiently investigated in the context of the turbulence cascade. 

The concept of microscopically reversible turbulence has been treated in the literature.
\citet{gallavotti1997dynamical} proposed a reversible formulation of dissipation in an
attempt to apply the theory of hyperbolic dynamical systems to turbulence, and to study
fluctuations out of equilibrium. This idea was later applied by \citet{biferale1998time} to
study a reversible shell model of the turbulence cascade under weak departures from
equilibrium. \citet{rondoni1999fluctuations} and \citet{gallavotti2004lyapunov} further
extended reversible models to two-dimensional turbulence, proving that reversible
dissipative systems can properly represent some aspects of turbulence dynamics.
Reversibility is also a property of some common large-eddy simulation models, which
reproduce the odd symmetry of the energy fluxes on the velocities and yield fully reversible
representations of the sub-grid stresses \citep{bardina1980improved, germano1991dynamic,
winckelmans2001explicit}. Despite the statistical irreversibility of the energy cascade,
previous investigations have also shown that it is possible to reverse turbulence in time.
\citet{carati2001modelling} constructed a reversible system using a standard dynamic
Smagorinsky model from which molecular dissipation has been removed. When reversed in time after decaying for
a while, this system recovers all its lost energy and other turbulent quantities; during the 
inverse evolution, the system develops a sustained inverse energy cascade towards the large
scales. This \refb{numerical experiment} shows that, even if we empirically know that spontaneously observing
an extended inverse cascade is extremely unlikely, such cascades are possible, thus exposing
their probabilistic nature. %entropic nature.
We exploit here the experimental system in \citet{carati2001modelling} as a tool to explore
the statistical irreversibility of the energy cascade.

As originally argued by \citet{loshmidt}, the analysis of the equations of motion of any
reversible dynamical system reveals that for each direct evolution there exists a
dual inverse evolution, leading to the paradox of how macroscopically irreversible
dynamics stems from microscopically reversible dynamics. This apparent inconsistency is
bridged with the concepts of entropy and entropy production, which account for the disparate
probabilities of direct and inverse evolutions. 
The concept of entropy is supported on the description of physical systems by time-invariant probability distributions in a highly-dimensional phase space. 
While this approach has been successfully applied to describe equilibrium systems, its application to out-of-equilibrium systems is still incomplete.

\refc{In the case of turbulence, the application of entropy to justify the direction of the cascade dates back several decades. 
\citet{kraichnan1967inertial} used absolute equilibrium ensembles of the inviscid Navier--Stokes equations to predict the inverse cascade of energy in 2D turbulence.
\refm{These equilibrium ensembles are derived from the equipartition distribution, which maximises a Gibbs entropy, implicitly suggesting
a quantitative connection between entropy and energy fluxes.}
Although this approach has proved successful to predict or justify the direction of fluxes in diverse out-of-equilibrium turbulent systems, including magnetohydrodynamic turbulence \citep{frisch1975possibility} and incompressible 3D turbulence \citep{orszag1974lectures}, it fails to offer useful information on the mechanisms that cause the prevalence of a particular direction of the fluxes. 
For that purpose, rather than time-invariant equilibrium distributions,
we must consider the dynamical information encoded in the temporal evolution of out-of-equilibrium probability distributions towards absolute equilibrium.}
Unfortunately, turbulence exhibits a wide range of spatial and temporal scales, which renders its description in a highly-dimensional
phase-space all but intractable. 

% \refc{In the case of turbulence, the idea of an entropy  Kraichnan \citep{}}
% Perhaps the closest approach to a `second law' for dynamical systems out
% of equilibrium are the fluctuation relations (FR) of \cite{evans2002fluctuation}, which
% quantify entropy production as phase-space volume contraction. Although the FR agree
% with the observational evidence, and might be applied to reversible LES of turbulence,. These limitations
% stem from two aspects of the FR. First, they treat entropy production as strictly connected to the
% dissipative mechanisms of the flow, but not to the inertial aspects of the energy cascade.
% Second, the fluctuation relations are global in nature and disregard the spatial characteristics of turbulent flows.
%
% The limitations of the FR are common to many of the tools used in dynamical system theory.

The main contribution of this work is to bridge the gap between the evolution of probability distributions in phase space,
and the physical structure of turbulent flows. We characterise the energy cascade as a local process in physical space, which can be robustly quantified independently of the definition of energy fluxes.
These results extend previous works on the locality of the energy cascade 
\citep{meneveau1994lagrangian,eyink2005locality,domaradzki2007analysis,eyink2009localness,cardesa2015temporal,cardesa2017turbulent,doan2018scale}. 
We consider the statistics of local energy transfer events to study the probability of inverse cascades over restricted domains, 
connecting \refm{statistical irreversibility, a property of phase-space ensembles related to entropy production,
with the local dynamics of turbulence in physical space.}
%entropy production as a macroscopic quantity with the statistics of local dynamics in physical space.

%The entropic nature of the cascade requires that we characterise its mechanisms
%from a probabilistic perspective. 
The probabilistic nature of the cascade requires that we characterise its mechanism also from a
probabilistic perspective.
A meaningful description of the energy cascade must first
identify the dynamically relevant mechanisms related to energy transfer, and subsequently
establish the causes of the prevalence of the direct mechanisms over the inverse ones.
We address both by analysing the filtered velocity gradient tensor through
its invariants \citep{naso2005scale,lozano2016multiscale,danish2018multiscale},
and by determining their relation to the local energy fluxes in physical space. 
These invariants describe the local geometry of turbulent flows, and offer a compact 
representation of the structure and dynamics of the velocity gradients. 
We will show that the structure of the direct cascade differs
substantially from that of the inverse cascade, and that differences are most significant 
in regions where the dynamics of the rate-of-strain tensor dominates over the vorticity vector. 
Moreover, we show that strain-dominated regions are 
responsible for most of the local energy fluxes, 
evidencing the relevant role of the rate-of-strain tensor in the energy cascade
and in the statistical irreversibility turbulent flows.

Following these results we propose an \refm{probabilistic} argument to explain
the prevalence of direct energy transfer by taking into consideration the interaction of the
rate-of-strain tensor with the non-local component of the pressure Hessian. 
In this frame, we justify the higher probability of direct
over inverse cascades by noting that the latter require the organisation of a 
large number of spatial degrees of freedom, whereas the direct cascade results from space-local dynamics.

%%%%%%%%%%%
This paper is organised as follows. 
\refm{In $\S$\ref{sec:entropy}, we review the origin of statistical irreversibility in the turbulence cascade as explained by 
the evolution of out-of-equilibrium ensembles and the production of information-entropy.}
In $\S$\ref{sec:model}, we present the reversible sub-grid model and the experimental set-up for the reversible turbulent system that constitutes the foundations of this investigation.
In order to examine the entropic \refm{(probabilistic)} nature of the turbulence cascade,
we conduct different \refb{numerical experiments} on this system, which are detailed in $\S$\ref{sec:num}.
In $\S$\ref{sec:prob} we characterise the distribution of inverse trajectories in phase space and the
structure of the energy cascade in physical space.
In $\S$\ref{sec:qr}, we compare the structure and dynamics of the direct and inverse evolutions through the invariants of the velocity gradient tensor 
and their relation to the local energy fluxes. Finally, conclusions are offered in $\S$\ref{sec:conc}. 

%---------------------------------------------------------------------------------------
\section{Entropy production and the turbulence cascade}
\label{sec:entropy}

\refm{We explore in this section the origin of irreversibility in the inertial energy cascade from the perspective
of statistical mechanics. We use the concept of entropy and entropy production to justify the prevalence of direct energy cascades,
and suggest a connection between these quantities and the interscale energy fluxes.}
\refc{Previous papers in this direction are \citet{orszag1974lectures} and \citet{holloway1986eddies}.}

Let us consider an $n$-dimensional state vector in phase space,
$\boldsymbol{\chi}=(\chi_1,\chi_2,\dots,\chi_n)$, representing the $n$ degrees of freedom of
a deterministic dynamical system, and its probability density, $P(\boldsymbol{\chi})$.
Assuming that this representation of the system satisfies the Liouville theorem, i.e. that
the dynamics preserve phase-space volume and therefore probability density along trajectories, we partition
the accessible phase space in $\gamma=1, 2,\dots, m$ coarse-grained subsets of volume
$\omega_\gamma$. The probability of finding a state in $\gamma$ is
\begin{equation}
P_\gamma=\int_{\omega_\gamma} P(\boldsymbol{\chi}) \dd \boldsymbol{\chi},
\end{equation}
where the integral is taken over $\omega_\gamma$. We define a Gibbs coarse-grained
entropy,
\begin{equation}
H=-\sum_{\gamma} {P_\gamma}\log\Big(P_\gamma\frac{\Upsilon}{\omega_\gamma}\Big),
\label{equ:entrop}
\end{equation}
where $\Upsilon=\sum_{\gamma}\omega_\gamma$, and the summation is done over all the
sub-volumes of the partition. This entropy is maximised for the coarse-grained probability
distribution $P_\gamma/\omega_\gamma=\text{const}$, %, which is satisfied by a uniform $P(\boldsymbol{\chi})$. 
although the value of this maximum depends on the geometrical properties
of the partition, such as the volume of the subsets, and can only be defined up to a
constant.
%\r{equ:entrop} is always maximised by this distribution. 
% Moreover, \r{equ:entrop} is proportional to the logarithm of the probability of a given ensemble of states, so that,
% given enough time to achieve equilibrium, the most probable state of a conservative system
% with mixing chaotic dynamics is the one that maximises $H$. 
This definition of entropy can also be connected
with information theory, such that the information on an ensemble of realisations is proportional to
minus its entropy \citep{latora1999kolmogorov}.

Let us consider the Euler equations projected on a truncated Fourier basis, $k\le
k_{max}$, where $k$ is the wave-number magnitude. This system conserves energy and satisfies
Liouville's equation for the Fourier coefficients of the fluid velocity
\citep{orszag1974lectures, kraichnan1975remarks}. If we choose an ensemble of states far
from equilibrium, such as a set of velocity fields with a given total kinetic energy and an
energy spectrum proportional to $k^{-5/3}$ \citep{kolmogorov1941local}, $H$ is initially low
because the states are localised in a special subset of phase space. 
As the ensemble evolves, the probability density is conserved along each trajectory,
$P(\boldsymbol{\chi}(t))=\text{const}$, and the chaotic interaction of a large number of degrees of freedom
leads to the phase-space `mixing' of the probability density,
resulting in the homogenisation of the coarse-grained probabilities, $P_\gamma$.
As a consequence, $H$ increases until the coarse-grained probability over each
element of the partition reaches $P_\gamma/\omega_\gamma=\text{const}$, which maximises
\r{equ:entrop} and corresponds to absolute equilibrium. 
\refm{The evolution of $H$ towards a maximum manifests the second law of
thermodynamics, and explains the emergence of statistical irreversibility in microscopically reversible systems out of equilibrium.} 

% In the phase space of the Fourier coefficients, this equilibrium is represented by a
% statistically uniform coarse-grained probability distribution of the velocity intensities across all modes,
% so that the equilibrium energy spectrum is proportional to the number of modes in each wavenumber shell, $4\pi k^2$.

\refc{In the phase space representation in terms of the Fourier coefficients, mixing is implemented by the conservative exchange of energy among triads of Fourier modes,
which distributes energy evenly across all modes \citep{kraichnan1959structure}}, so that the equilibrium energy spectrum is proportional to the number of modes in each wavenumber shell, $4\pi k^2$.
This is the spectrum of the most probable macroscopic state, i.e. the spectrum that represents the largest
number of microscopic realisations. States in the equilibrium ensemble have more energy in the small scales than
the states in our initial out-of-equilibrium ensemble, implying that the evolution
towards equilibrium and the increase of entropy corresponds, on average, 
to energy flux towards the small scales.
However, note that the scale is not a property of the phase space itself, nor of the
entropy. In our example, the scale has been overlaid on the system by labelling the Fourier
coefficients by their wavenumbers, where a greater number of coefficients are tagged as `small scale' than as `large scale'.
Moreover, the evolution towards equilibrium is true only in a statistical sense. 
Since there is no deterministic `force' driving the system to equilibrium, 
it is possible to find trajectories in the ensemble 
for which energy sloshes back and forth among different scales. 
One cascade direction is simply more probable than the other.
% Notationally, we define a cascade as `direct' 
% if it goes in the most probable direction, and `inverse' otherwise.

In the previous discussion we have disregarded the dynamics of the flow, whose
evolution has been reduced to the chaotic mixing in phase space. In reality, the system is
restricted by the equations of motion, and not all evolutions are possible.
At the very least, the system only evolves on the hypersurfaces defined by
its invariants, % which may have different mode counts than the full space, and 
which thus determine different equilibrium spectra. The most obvious invariant of the inviscid
Euler equations is the total energy, which can be written as $\mathcal{E}=\sum_k \chi_k^2$.
But, because this formula does not explicitly involve the scale, the restriction to an
energy shell does not modify energy equipartition, nor the $k^2$ spectrum.
When there are more than one invariant, the accessible phase space is the intersection
of their corresponding isosurfaces, which generically depends on the relative magnitude of the total
conserved quantities. This is the case of the two-dimensional inviscid Euler equations, which also conserve the enstrophy \citep{kraichnan1980two},
or the inviscid and infinitely conducting magnetohydrodynamic equations, which also conserve the magnetic helicity \citep{frisch1975possibility}.

Unfortunately, although the form of the equilibrium spectrum suggests where the system
`would like to go', it does not by itself determine the direction of the cascades in
out-of-equilibrium steady states. These states can be established by imposing boundary
conditions in phase space, such as forcing the large scales, and adding dissipation at the
small ones, resulting in a steady flux of energy from one boundary to the
other. The intuition is that the cascade develops as the
energy moves among scales, trying to establish the equilibrium spectrum, but that this final
state is never achieved because the boundary conditions sustain the energy flux and maintain
the system out of equilibrium.

In summary, \refm{statistical irreversibility and the direct energy cascade} arise in out-of-equilibrium
ensembles of the truncated Euler equations as consequences of three aspects of inertial
dynamics: the first one is the conservation of phase-space probabilities along trajectories,
due to Liouville; the second one is the strongly mixing nature of turbulent dynamics in the
inertial range; and the third one is that small scales are represented by a much higher
number of degrees of freedom than larger ones. These properties intuitively answer the
central question of why direct cascades are more probable than inverse ones. Although
possible, the latter are inconsistent with the tendency of turbulent flows to evolve towards
equilibrium, because they take energy from the more numerous small eddies and `organise' it
into less common larger ones.

However, %as we noted for our model sand pile, 
these general statistical concepts say
little about the dynamics of the cascade. 
\refm{The main contribution of our work is to explain the origin of statistical irreversibility 
by considering the structure of the energy cascade in physical space, 
connecting the description of turbulence as a dynamical system evolving in a highly-dimensional phase-space
to the space-local physical structure of isotropic turbulence}. 
We do so by describing the physical mechanisms that locally determine the prevalence of the direct over the inverse energy cascade.

% In the following sections, we will demonstrate that positive entropy production, as represented by the tendency , stems from the probability distribution of local energy-transfer events in physical space.
% We find that these events are related to particular local geometrical properties of the flow, as described by the invariants of the velocity gradient tensor. 
% The dynamical analysis of these geometrical properties allows us to give a physical explanation for the prevalence of direct cascades over inverse cascades.

% This analysis bridges the gap between the powerful but intractable concept of phase space and the fundamental structure of turbulent flows.
% We relate the conservative, chaotic-mixing properties of out-of-equilibrium ensembles in phase space to local dynamics in physical space, identifying the mechanisms of entropy production in turbulent flows.

% --------------------------------------------------------------------------
% \newpage
% --------------------------------------------------------------------------
\section{The reversible sub-grid model}
\label{sec:model}

A popular technique to alleviate the huge computational cost of simulating
industrial turbulent flows is large-eddy simulation, which filters out the flow scales below
a prescribed cutoff length, and only retains the dynamics of the larger eddies. We use it
here to generate microscopically reversible turbulence. The equations governing the large
scales are obtained by filtering the incompressible Navier--Stokes equations,
\begin{equation}
\begin{aligned}
\partial_t \overline  u_i+\overline u_j\partial_j \overline  u_i&=-\partial_i \overline p +\partial_j \tau_{ij},\\
\partial_i \overline u_i&=0,
\end{aligned}
\label{eq:0}
\end{equation}
where the overline $(\overline \cdot)$ represents filtering at the cutoff length $\overline
\Delta$, $u_i$ is the $i$-th component of the velocity vector $\boldsymbol u=(u_i)$, with
$i=1\ldots 3$, $\partial_i$ is the partial derivative with respect to the $i$-th direction,
$p$ is a modified pressure, and repeated indices imply summation. We assume that the cutoff
length is much larger than the viscous scale, and neglect in \r{eq:0} the effect of
viscosity on the resolved scales. The interaction of the scales below the cutoff
filter with the resolved ones is represented by the sub-grid stress (SGS) tensor,
$\tau_{ij}=\overline{u}_i \overline{u}_j-\overline{u_iu_j}$, which is unknown and must be
modelled.

One of the consequences of this interaction is an energy flux towards or from the unresolved
scales, which derives from triple products of the velocity field and its derivatives, and
is implicitly time-reversible. However, this flux is often modelled as a dissipative
energy sink, destroying time reversibility and the possibility of inverse energy fluxes
(backscatter).
In an attempt to yield a more realistic representation of the dynamics of the energy
cascade, some SGS models try to reproduce backscatter, which also restores 
the time-reversibility of energy fluxes.
This is the case for the family of dynamic models, which
are designed to adapt the effect of the unresolved scales to the state of the resolved flow,
eliminating tuning parameters. A widely used reversible model is the dynamic Smagorinsky
model of \cite{germano1991dynamic}, based upon the assumption that the cutoff filter
lies within the self-similar inertial range of scales, and that the sub-grid stresses at the
filter scale can be matched locally to those at a coarser test filter. This idea is applied
to the classical \citet{smagorinsky1963general} model in which sub-grid stresses are assumed
to be parallel to the rate-of-strain tensor of the resolved scales,
$\overline{S}_{ij}=\tfrac{1}{2}(\partial_i \overline{u}_j +\partial_j \overline{u}_i)$, such that
$\tau_{ij}^T=2 \nu_s \overline{S}_{ij}$, where $\nu_s$ is referred to as the eddy
viscosity, and the `$T$' superscript refers to the traceless part of the tensor. Introducing
$\overline \Delta$ and $|\overline{S}|$ as characteristic length and time scales,
respectively, and a dimensionless scalar parameter $C$, the model is
\begin{equation}
\tau_{ij}^T = 2C \overline{ \Delta}^2 |\overline{S}| \overline{S}_{ij},
\label{equ:fluxes}
\end{equation}
where $\nu_s=C \overline{ \Delta}^2 |\overline{S}|$, and $|S|=\sqrt{2 S_{lm}S_{lm}}$.
Filtering \r{eq:0} with a test filter of width $\widetilde \Delta=2\overline \Delta$,
denoted by $(\tilde \cdot)$, we obtain expressions for the sub-grid stresses at both
scales, which are matched to obtain an equation for $C$,
\begin{equation}
%C_s \widetilde \Delta |\widetilde{\overline{S}}| \widetilde{\overline{S}} =\widetilde{\overline{u} \overline{u}} - \widetilde{\overline{u}} \widetilde{\overline{u}}+C_s \overline{\Delta}^2 \widetilde{|\overline{S}|\overline{S}}.
C \mathcal{M}_{ij}+\mathcal{L}^T_{ij}=0,
\label{equ:germano}
\end{equation}
where $\mathcal{M}_{ij}= {\widetilde{\Delta}}^2 |\widetilde{\overline{S}}|
\widetilde{\overline{S}}_{ij}- {\overline{\Delta}}^2\widetilde{|\overline{S}|\overline{S}_{ij}}$, and  
$\mathcal{L}_{ij}=\tfrac{1}{2} (\widetilde{\overline{u}_i\overline{u}_j} -
\widetilde{\overline{u}}_i \widetilde{\overline{u}}_j)$. A spatially local least-square solution,
$C_\ell$, is obtained by contracting \r{equ:germano} with
$\mathcal{M}_{ij}$,
\begin{equation}
 C_\ell(\boldsymbol x, t;\overline\Delta)=\frac{\mathcal{L}^T_{ij}\mathcal{M}_{ij} }{ \mathcal{M}_{ij}\mathcal{M}_{ij}}.
 \label{equ:germano1}
\end{equation}
This formulation occasionally produces local negative dissipation, $C_\ell(\boldsymbol x,
t)<0$, which may lead to undesirable numerical instabilities
\citep{ghosal1995dynamic,MenLundCab1996dynamic}. To avoid this problem, \r{equ:germano} is
often spatially averaged after contraction to obtain a mean value for the dynamic parameter
\citep{lilly1992proposed},
\begin{equation}
 C_s(t; \overline\Delta)=\frac{ \langle \mathcal{L}^T_{ij}
 \mathcal{M}_{ij} \rangle}{ \langle \mathcal{M}_{ij}\mathcal{M}_{ij} \rangle},
 \label{equ:germano2}
\end{equation}
where $\langle\cdot \rangle$ denotes spatial averaging over the computational box.
\refc{Since $\tau_{ij} {\overline S}_{ij}= C_s {\overline \Delta}^2 |{\overline S}_{ij}|^3$},
a positive $C_s$ implies that energy flows from the resolved to the unresolved scales,
and defines a direct cascade. The sign of $C_s$ depends on the velocity field through
$\mathcal{M}_{ij}$, which is odd in the rate-of-strain tensor, so that
$C_s(-\boldsymbol{u})=-C_s(\boldsymbol{u})$. Given a flow field with $C_s>0$, a change in
the sign of the velocities leads to another one with negative eddy viscosity.
%%

% ===========================================
\begin{figure}
  \begin{center}
    \includegraphics[width=0.45\textwidth]{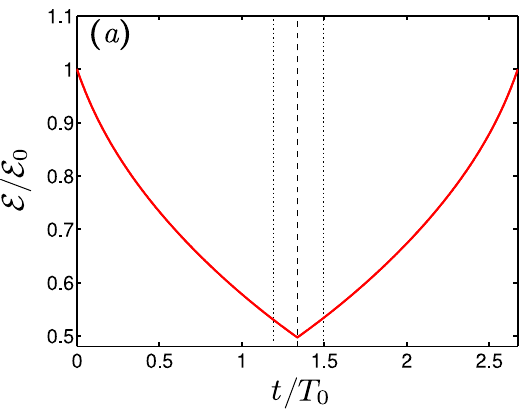}
    \includegraphics[width=0.49\textwidth]{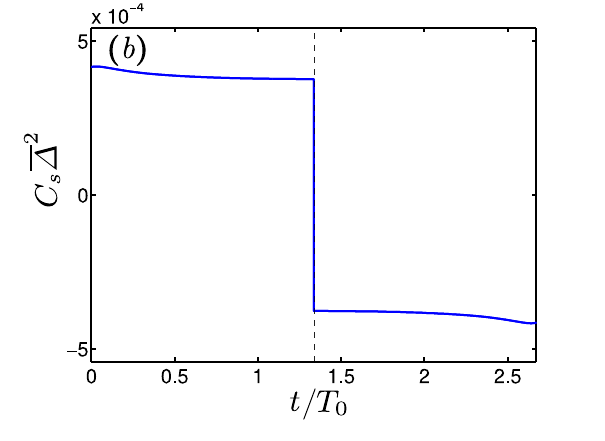}

    \includegraphics[width=0.45\textwidth]{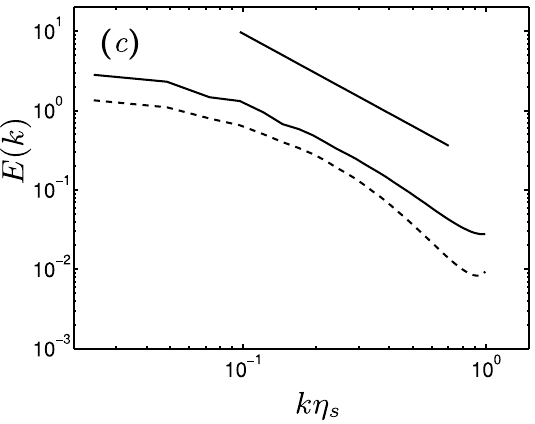}
    \includegraphics[width=0.49\textwidth]{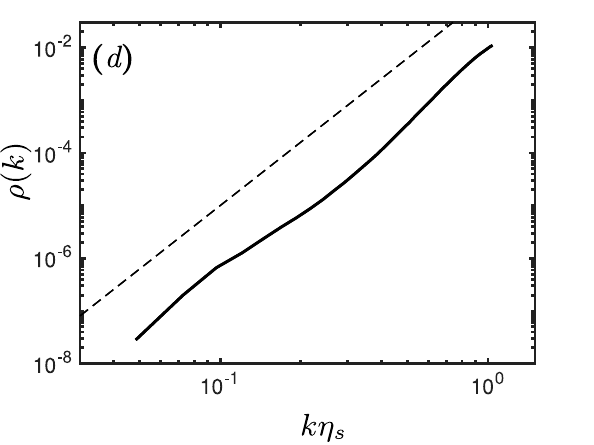}
\caption{Evolution of: (\aaa) mean kinetic energy, normalised with the initial energy
$\mathcal{E}_0$, and (\bbb) $C_s\overline \Delta^2$ as a function of time (arbitrary units).
The dashed line in (\aaa,\bbb) is $t=t_{inv}$, and the dotted lines in (\aaa) are
$t=t_{stats}$, used in \S\ref{sec:qr}.
(\ccc) Energy spectrum $E(k)$. \solid, $t=0$ and $t=2t_{inv}$; \dashed, $t=t_{inv}$.
The diagonal line is $E(k)\propto k^{-5/3}$.
(\ddd) Spectrum of the error between the velocity fields at $t=2t_{inv}$ and
$t=0$, as a function of wavenumber. Dashed line denotes $k^4$.}
%    
%\includegraphics[width=0.5\textwidth]{Cs_1.pdf}\includegraphics[width=0.5\textwidth]{Cs_2.pdf}
      %\includegraphics[width=0.5\textwidth]{diss_1.pdf}\includegraphics[width=0.5\textwidth]{diss_2.pdf}
   
%   \caption{Energy and $C_s$ evolution as a function of time in arbitrary units. }
    \label{fig:parameters_evolution}
  \end{center}
\end{figure}
% ===========================================

This property allows us to construct a reversible turbulent system using the dynamic
Smagorinsky model and removing molecular viscosity. The equations for the resolved velocity
field, $\overline{\boldsymbol{u}}$, are
\begin{equation}
\begin{aligned}
  \partial_t u_i+u_j\partial_j u_i&=-\partial_i p + C_s \overline{\Delta}^2 \partial_j {|S|} S_{ij},\\
  \partial_i u_i&=0,
\end{aligned}
\label{eq:1}
\end{equation}
together with (\ref{equ:germano2}), where the filtering notation has been dropped for
conciseness. The equations \r{eq:1} are invariant to the transformation $t\rightarrow -t$ and
$\boldsymbol{u} \rightarrow -\boldsymbol{u}$, and changing the sign of the velocities is
equivalent to inverting the time axis \refc{\citep{carati2001modelling}}.

%%%%%%%%%%%%%%%%%%%%%%%%%%%%%%%%%%%%%%%%%%%%%%%
\section{Experiments on turbulence under time reversal}
\label{sec:num} \label{sec:exp}

% --------------------------------------------------------------------------------------------
\subsection{Numerical setup}

A standard \citet{rogallo1981numerical} code is used to perform a series of experiments on
decaying triply-periodic homogeneous turbulence in a $(2\pi)^3$ box. Equations \r{eq:1} are
solved with the dynamic Smagorinsky model described in the previous section. The three
velocity components are projected on a Fourier basis, and the non-linear terms are
calculated using a fully dealiased pseudo-spectral method with a 2/3 truncation rule
\citep{canuto2012spectral}. We denote the wave-number vector by $\boldsymbol{k}$, and its
magnitude by $k=|{\boldsymbol{k}}|$. The number of physical points in each direction before
dealiasing is $N=128$, so that the highest fully resolved wave number is $k_{max}=42$. An
explicit third-order Runge-Kutta is used for temporal integration, and the time
step is adjusted to constant Courant-Friedrichs-Lewy number equal to $0.2$.
A Gaussian filter whose Fourier expression is
\begin{equation}
G(k;\Delta)=\exp(-\Delta^2k^2/24),
\label{eq:gfil}
\end{equation}
where $\Delta$ is the filter width \citep{aoyama2005statistics}, is used to evaluate $C_s$
in \r{equ:germano2}. The equations are explicitly filtered at the cutoff filter
$\overline{\Delta}$, and at the test filter $\widetilde
\Delta=2\overline{\Delta}=\Delta_{g}\sqrt{6}$, where $\Delta_{g}=2\pi/N$ is the
grid resolution before dealiasing. Although the explicit cutoff filter is not strictly
necessary, it is used for consistency and numerical stability.

The energy spectrum is defined as
\begin{equation}
E(k;\boldsymbol u)=2\pi k^2\langle \widehat{u}_i \widehat{u}_i^*\rangle_k,
\end{equation}
where $\widehat{u}_i(\boldsymbol{k})$ are the Fourier coefficients of the $i$-th velocity
component, the asterisk is complex conjugation, and $\langle \cdot \rangle_k$ denotes
averaging over shells of thickness $k\pm 0.5$. The kinetic energy per unit mass is
$\mathcal{E}=\tfrac{1}{2}\langle {u_i u_i}\rangle=\sum_k E(k,t)$, where $\sum_k$ denotes
summation over all wave-numbers.

Meaningful units are required to characterise the simulations. The standard
reference for the small scales in Navier--Stokes turbulence are Kolmogorov units, but they are not
applicable here because of the absence of molecular viscosity. Instead we derive
pseudo-Kolmogorov units using the mean eddy viscosity, $\langle\nu_s\rangle= C_s\overline
{\Delta}^2 \langle |S| \rangle$, and the mean sub-grid energy transfer at the cutoff scale,
$\langle\epsilon_s\rangle= C_s\overline{\Delta}^2 \langle |S|^3 \rangle$.
The time and length scale derived from these quantities are
$\tau_s=(\langle\nu_s\rangle/\langle \epsilon_s \rangle)^{1/2}$ and
$\eta_s=(\langle\nu\rangle^3_s/\langle \epsilon_s \rangle)^{1/4}$, respectively.
We find these units to be consistent with the physics of turbulence, such that the peak 
of the enstrophy spectrum, $k^2E(k)$, is located at $\sim25\eta_s$. 
In DNSs, this peak lies at $\sim20\eta$.

% The dissipation peak, which is identified as the maxima of
% \begin{equation}
% 4\pi k^2\mathrm{Re}\langle \widehat{u}^*_i k_j\widehat{|S|S_{ij}}\rangle_k,
% \end{equation}
% is located at the cutoff wavenumber and is consistent with $\eta_s\sim0.5\Delta_g$, which is smaller than the resolution grid.
% \jj{Creo recordar que hab\'ia razones para creerse que esas unidades ten\'ian sentido. E.g.
% Donde est\'a el pico de disipaci\'on w.r.t. $\eta_s$?}
% 
% \av{El pico de la dissipaci\'on esta en el \'ultimo modo, pero $\eta_s$ es m?s peque?o 
% que el ancho del filtro, que es lo que en su d\'a decidimos que tenia sentido.
% Lo que pasa es que $\nu_s$ baja mucho en la simulacion y hace que suba el $Re_L$,
% lo cual no tiene mucho sentido cuando la relaci?n $L/\eta_s$ se mantiene casi 
% igual porque baja un poco $L$ y un poco $\eta_s$.
% }
% 

The larger eddies are characterised by the integral length, velocity and \refc{time} scales
\citep{bat53}, respectively defined as
\begin{equation}
L=\pi \frac{ \sum_k^{} E(k,t)/k}{ \sum_k^{} E(k,t)},
\end{equation}
$u'=\sqrt{2\mathcal{E}/3}$ \refc{and $T=L/u'$}. The Reynolds number based on the
integral scale is defined as $Re_L=u' L/\langle\nu_s\rangle$, and the separation of scales
in the simulation is represented by the ratio $L/\eta_s$.

In the following we study the energy flux both in Fourier space, averaged over the
computational box, and locally in physical space.

The energy flux across the surface of a sphere in Fourier space with wave-number magnitude
$k$ can be expressed as
\begin{equation}
\begin{aligned}
\Pi(k)&=\sum_{q<k} 4\pi q^2\mathrm{Re}\langle \widehat{u}^*_i(\boldsymbol{q})\widehat{\mathcal{D}}_i(\boldsymbol{q})\rangle_q, 
\label{eq:euler_2}
\end{aligned}
\end{equation}
where $\mathcal{D}_i=u_j\partial_j u_i$, and a negative $\Pi$ denotes energy flowing to larger scales. 

We use two different definitions of the local inter-scale energy flux in physical space,
\begin{equation}
\begin{aligned}
\Sigma(\boldsymbol x,t; \Delta) &=  \tau_{ij}{S}_{ij}, \\
\Psi(\boldsymbol x,t; \Delta)   &= -u_i\partial_j \tau_{ij},
\end{aligned}
\label{eq:sigma}
\end{equation} 
where the velocity field, the rate-of-strain tensor, and $\tau_{ij}$ are
filtered with \r{eq:gfil} at scale $\Delta$. Both are standard quantities in the analysis of
the turbulence cascade \citep{ meneveau1994lagrangian,borue1998local, aoyama2005statistics,
cardesa2015temporal}, and are related by $\Psi=\Sigma - \partial_j (u_i\tau_{ij})$. The
second term in the right-hand side of this relation is the divergence of an energy flux in
physical space, which has zero mean over the computational box, so that $\langle
\Sigma\rangle=\langle\Psi\rangle$. Positive values of $\Sigma$ and $\Psi$ indicate that the
energy is transferred towards the small scales.

Because the decaying system under study is statistically unsteady, averaging over an
ensemble of many realisations is required to extract time-dependent statistics. This is
generated using the following procedure. All the flow fields in the ensemble share an
initial energy spectrum, derived from a forced statistically-stationary simulation, and an
initial kinetic energy per unit mass. Each field is prepared by randomising the phases
of $\widehat{u}_i$, respecting continuity, and integrated for a fixed time $t_{start}$ up to
the $t_0=0$, where a fully turbulent state is deemed to have been reached, the experiment
begins, and statistics start to be compiled. 
The initial transient, $t_{start}$, common to
all the elements in the ensemble, is chosen so that $t_0$ is beyond the time at which
dissipation reaches its maximum, and the turbulent structure of the flow is fully developed.
\refc{Quantities at $t_0$ are denoted by a `0' subindex.}
Following this procedure, we generate an ensemble of $N_s=2000$ realisations, which are
evolved on graphic processing units (GPUs). 
\refc{The statistics below are compiled over all the members of this ensemble.}

It is found that both the small- and large-scale reference quantities defined above vary
little across an ensemble prepared in this way, with a standard deviation of the order of
$5\%$ with respect to their mean.

% --------------------------------------------------------------------------------------------
\subsection{Turbulence with a reversible model}\label{sec:carati}

The basic \refb{numerical experiment} is conducted as in \cite{carati2001modelling}. 
\refc{After preparing the ensemble of turbulent fields at $t=t_0=0$},
they are evolved for a fixed time $t_{inv}$, during which
some of the initial energy is exported by the model to the unresolved scales.
The simulations are then stopped and the sign of the three components of velocity reversed,
$\boldsymbol{u}\rightarrow -\boldsymbol{u}$. The new flow fields are used as the initial
conditions for the second part of the run from $t_{inv}$ to $2t_{inv}$, 
during which the flows evolve back to their original state,
recovering their original energy and the value of other turbulent quantities.
 
It is found that $\eta_s$ only changes by $5\%$ during the decay of the flow, while $\tau_s$
increases by a factor of $1.5$ from $t=0$ to $t_{inv}$. On the other hand, the large-scale
quantities, $L$ and $u'$ vary substantially as the flow decays. The main parameters
of the simulations, and the relation between large- and small-scale quantities, are given in
table \ref{tab:kd}.

% ================================================================
\begin{table}
  \begin{center}
\def~{\hphantom{0}}
  \begin{tabular}{lcccccccccc}
       & $N$  & $k_{max}$  &  $Re_{L}$ & $k_{max}\eta_s$  & $\Delta_{g}/\eta_s$ &  $t_{inv}/T_0$  & $L/\eta_s$ & $L/\Delta_{g}$ & $\epsilon_{s} L /u'^3$ & $N_s$  \\[3pt]
      $t=0$       & 128  &  42   &   214    &  0.94  & 2.2   &   1.3      &       63.3  & 28.3 & 1.3 & 2000  \\[3pt]
      $t=t_{inv}$ & & & 248 & 0.87 & 2.3 & & 63.1 & 26.1 & 1.0 & \\
  \end{tabular}
  \caption{Main parameters of the simulation averaged over the complete ensemble at time $t_0=0$ and $t_{inv}$. See the text for definitions.}
  \label{tab:kd}
  \end{center}
\end{table}
% ==========================================================

Figure \ref{fig:parameters_evolution}(\aaa, \bbb) presents the evolution of the kinetic
energy $\mathcal{E}$ and of $C_s$ as a function of time for a representative experiment. We
observe a clear symmetry with respect to $t_{inv}$ in both quantities, and we have checked
that this symmetry also holds for other turbulent quantities, such as the sub-grid energy
fluxes and the skewness of the velocity derivatives. Quantities that are odd with respect to
the velocity display odd symmetry in time with respect to $t_{inv}$, and even quantities
display even symmetry. In particular we expect $\boldsymbol u(2t_{inv})\simeq-\boldsymbol u(0)$

Figure \ref{fig:parameters_evolution}(\ccc) shows the energy spectrum $E(k)$ at times $t=0$,
$t_{inv}$ and $2t_{inv}$. Although there are no observable differences between the initial
and final spectra, we quantify this difference in scale using the spectrum of the error
between the velocity field at $t=0$ and minus the velocity field at $2t_{inv}$,
\begin{equation}
\rho(k)=\frac{2E(k;\boldsymbol{u}(0)+\boldsymbol{u}(2t_{inv}))}{E(k;\boldsymbol{u}(0)) + E(k;\boldsymbol{u}(2t_{inv}))},
\end{equation}
which is presented in figure \ref{fig:parameters_evolution}(\ddd).
% %%
% \begin{equation}
% \rho(k;\boldsymbol{\psi},\boldsymbol{\zeta})=\frac{2r_{\boldsymbol{\psi\zeta}}(k)}{r_{\boldsymbol{\psi\psi}}(k) + r_{\boldsymbol{\zeta\zeta}}(k)}, 
% \end{equation}
% where,
% \begin{equation}
% r_{\boldsymbol{\psi\zeta}}(k)=4\pi k^2\mathrm{Re}\langle \widehat{\psi}_i \widehat{\zeta}_i^*\rangle_k,
% \end{equation}
% %
% which fulfills that
% %%
% \begin{equation}
% \frac{E(k;\boldsymbol{\psi}-\boldsymbol{\zeta})}{E(k;\boldsymbol{\psi}) + E(k;\boldsymbol{\zeta})}=
% 1 - \rho(k;\boldsymbol{\psi},\boldsymbol{\zeta}),
% \end{equation}
% %%
% such that $\rho(k)=1$ indicates that the spectrum of the difference between to fields is zero.
% % \jj{(Actually, this should be computed after subtracting the mean of both variables. As it is, it would not be zero even for completely unrelated variables, in which case it would just measure the standard deviation of each variable with respect to its mean.)}
% Results for
% $\rho(k;{\boldsymbol{u}}(t_0),-{\boldsymbol{u}}(2t_{inv}))$ averaged over the ensemble are
% presented in figure \ref{fig:parameters_evolution}(\ddd). 
The spectrum of the error is of the order of $0.01$ in the cutoff wavenumber and 
decreases as $k^4$ with the wavenumber,
confirming that the flow fields in the forward and backward evolution are similar,
except for the opposite sign and minor differences in the small scales.
This suggests that the energy cascade is microscopically reversible in the inertial scales.
Sustained reverse cascades are possible in the system, even if only direct ones are observed in
practice, supporting the conclusion that the one-directional turbulence cascade is an entropic (probabilistic) effect, unrelated to the presence of an energy sink at the small scales.

\refc{This conclusion is in agreement with the empirical evidence that the kinetic energy dissipation in turbulent flows is independent of the Reynolds number \citep{sreeni84}. 
This phenomenon, known as the dissipative anomaly, exposes the surrogate nature of kinetic energy dissipation, which is controlled by large-scale dynamics through the energy cascade process \citep{taylor,kolmogorov1941local}. 
In agreement with the dissipative anomaly, we will present evidence that the energy dissipation is a consequence, rather than a cause, of the energy cascade.}

%Menos es mas
% \section{Experiments on time-reversible turbulence}
% \label{sec:exp}
% ---------------------------------------------------------------------------------------
\subsection{The reverse cascade without model}\label{sec:nomod}

The validity of the conclusions in the previous section depends on the ability of the model
system to represent turbulence dynamics. In particular, since the object of our study is the
turbulence cascade rather than the SGS model, it is necessary to \refb{assess} whether the
reversibility properties of the system, and the presence of a sustained backwards energy
cascade, stem from the SGS model, or whether intrinsic turbulent mechanisms are involved. In
this subsection, we show that the model injects energy at the smallest resolved scales,
but that the energy travels back to the large scales due to inertial mechanisms. In the
next subsection, we further demonstrate that the inverse cascade can exist for some time
even for irreversible formulations of the subgrid model.

%
%=======================================================================
\begin{figure}
  \begin{center}
    \includegraphics[width=0.48\textwidth]{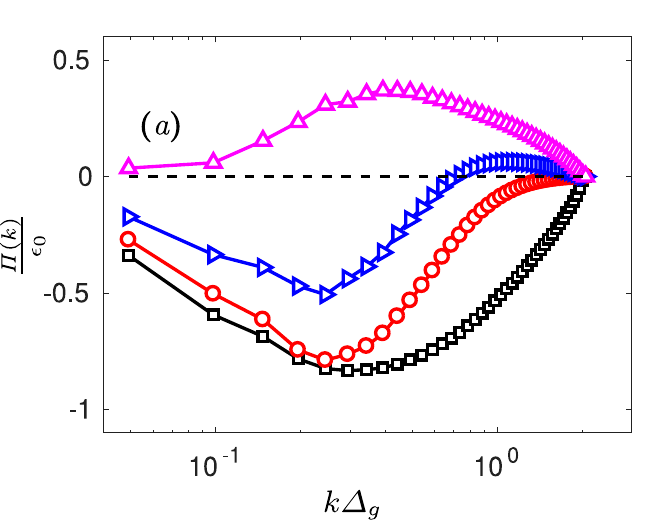}
    \includegraphics[width=0.48\textwidth]{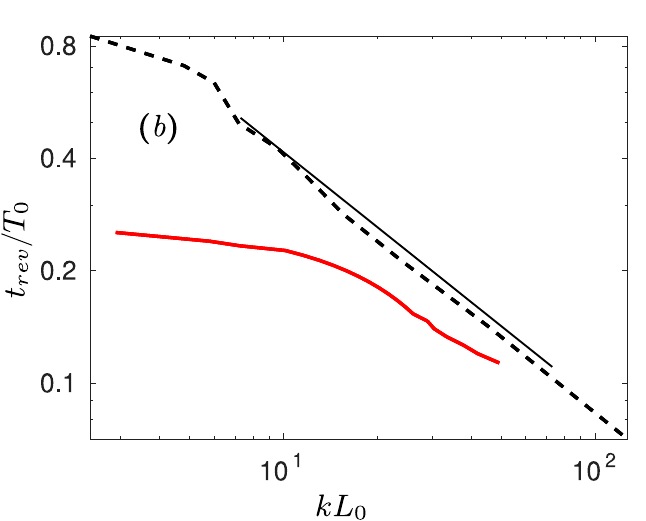}
      \caption{%
      (\aaa) Energy flux $\Pi(k)$ as a function of the time after removing the SGS model, \refc{$t/T_0$}:
      {\protect\linesquare{black}}, $0$;
       {\protect\linescircle{red}}, $0.1$;
       {\protect\linestriar{blue}}, $0.15$;
       {\protect\linestriau{magenta}}, $0.25$. 
%           {\protect\linestriau{red}}, $\Sigma$ at $5\Delta_{g}$;
%           {\protect\linesquare{blue}}, $\Sigma$ at $10\Delta_{g}$;
%           {\protect\linestriar{magenta}}, $C_\ell$ at $4.9\Delta_{g}$;
%           {\protect\linescircle{black}}, $\Sigma$ at $2.4\Delta_{g}$.
% 
      Energy flux normalised with $\epsilon_0={u'_0}^3/L_0$. 
      (\bbb) Reversal time $t_{rev}/T_0$ at which $\Pi=0$ as a function of wave-number $kL_0$. 
      Initial condition before the inversion of the velocities and the removal of the model obtained with: {\color{red}\solid}, reversible dynamic SGS model ($N=128$); {\dashed}, irreversible spectral SGS model ($N=512$). Solid black line corresponds to $t_{rev}/T_0=2{(kL_0)}^{-2/3}$.} 
    \label{fig:nomodel_1}
  \end{center}
\end{figure}
%====================================================================

If the SGS model contributed substantially to the presence of a sustained inverse cascade, 
its removal during the backward leg of the simulation would have an immediate
effect on the system, resulting in the instantaneous breakdown of the reverse cascade. This
is shown not to be the case by an experiment in which the model is removed at $t_{inv}$ by
letting $C_s=0$. The system then evolves according to the Euler equations
 \begin{equation}
  \begin{aligned}
  \partial_t u_i+u_j\partial_j u_i&=-\partial_i p, \\
  \partial_i u_i&=0,
  \end{aligned}
\label{eq:euler_1}
\end{equation}
which are conservative and include only inertial forces. The resulting evolution of the 
inertial energy flux across Fourier scales, $\Pi(k,t)$, defined in \r{eq:euler_2}, is
displayed in figure \ref{fig:nomodel_1}(\aaa) as a function of wavenumber and time.

At $t_{inv}$, energy fluxes are negative at all scales and energy flows towards the large
scales. Shortly after, at $t-t_{inv}=0.1T_0$, the inverse cascade begins to break down at the
small scales, while it continues to flow backwards across most wavenumbers. This continues
to be true at $t=0.15T_0$, where the direct cascade at the smallest scales coexists with a
reverse one at the larger ones. The wavenumber separating the two regimes moves
progressively towards larger scales, and the whole cascade becomes direct after $t\simeq
0.25T_0$.

Figure \ref{fig:nomodel_1}(\bbb) shows the dependence on the wavenumber of the
time $t_{rev}(k)$ at which $\Pi(k,t_{rev})=0$. It is significant that, for wavenumbers that
can be considered inertial, $kL_0\gg 1$, it follows $t_{rev}/T_0\approx 2(k L_0)^{-2/3}$, which
is consistent with the \citet{kolmogorov1941local} self-similar structure of the
cascade. 
If we assume a Kolmogorov spectrum, $E(k)=C_K \epsilon_s^{2/3} k^{-5/3}$, where $C_K\approx 1.5$ is the Kolmogorov constant \citep{pope2001turbulent}, 
and $\epsilon_s$ is the energy flux through wavenumber $k$, which is equal to the energy dissipation, the energy above a given wavenumber is
\begin{equation}
\mathcal{E}(k)=\int^{\infty}_{k} E(q)\mathrm{d}q \approx \frac{3}{2}C_K \epsilon_s^{2/3} k^{-2/3}.
\label{eq:ek} 
\end{equation}
When the SGS model is removed, an inverse cascade can only be sustained by drawing energy from this reservoir. 
Independently of whether the cascade can maintain its integrity from the informational point of view, 
the time it takes for a flux $-\epsilon_s$ to drain (\ref{eq:ek}) can be expressed as
\begin{equation}
\frac{t_{max}}{T_0}\approx \frac{3}{2} C_K \left( \frac{u'^3}{L_0\epsilon_s} \right)^{1/3} (k L_0)^{-2/3} 
\approx 2 (k L_0)^{-2/3},
\label{eq:333}
\end{equation}
where we have used values from table \ref{tab:kd2}. As shown in figure \ref{fig:nomodel_1}(b),
this approximation is only 50\% larger than the observed reversal time for the initial conditions 
prepared with the reversible SGS model.

% In essence, the reverse cascade survives while it has energy to draw from. 
%
% The time over which the information on the inverse cascade is destroyed is much longer, as shown by figure \ref{fig:parameters_evolution}(a),
% in which the cascade survives to essentially recover its forward initial conditions.

% \refc{These results show that, when the SGS model is removed, the inverse energy cascade 
% is destroyed in the large scales of the flow because of the propagation of strong disturbances across scales, 
% but not because the model is missing in those scales.}
The simplest conclusion is that the model just acts as a source to provide the small scales
with energy as they are being depleted by the flux towards larger eddies. If this
source is missing, the predominant forward cascade reappears, but the inertial mechanisms
are able to maintain the reverse cascade process as long as energy is available.

% We will show in $\S$\ref{sec:qr} that this delay is related to the reconstruction of the original geometry of the velocity gradients. 

% ---------------------------------------------------------------------------------------
\subsection{The effect of irreversible models}\label{sec:vismod}

We have repeated the same experiment for an irreversible spectral SGS model
\citep{metais1992spectral} with a higher resolution $(N=512)$. Results of $t_{rev}(k)$ for this experiment are shown by the dashed line in figure \ref{fig:nomodel_1}(\bbb).
% Despite the irreversibility of the SGS model used here,
The behaviour is similar to the experiment in the previous section, but the wider separation of
scales in this simulation allows the preservation of an inverse cascade for times of the
order of the integral time. This experiment confirms that the microscopic reversibility of the
inertial scales is independent of the type of dissipation,
and holds as long as the information of the direct cascade process is not destroyed 
by the LES model. This experiment suggests that microscopic reversibility 
should also hold in the inertial range of fully developed Navier--Stokes turbulence.
\refc{ Note that the good agreement of the reversal time with the energy estimate in (\ref{eq:333})
implies that $t_{rev}$ is very close to the maximum possible for a given energy.}

\section{Phase- and physical-space characterisation of the energy cascade}
\label{sec:prob}

% In this section we characterise the distribution of inverse trajectories in phase space, characterize the local nature of the energy cascade in physical space and 
% approaches to quantify the probability of these trajectories by analysing the local dynamics in physical space. 
% These approaches are connected with the theory of fluctuations and the evolution of infinitesimal perturbation in phase space. 

% We obtain qualitatively similar results from both the phase-space and physical-space analysis, which indicates the similarities of the 
% Both analysis yield qualitatively similar results, suggesting a connection 
% between 

\subsection{The geometry of phase space}
\label{sec:per}

\begin{figure}
  \begin{center}
    \includegraphics[width=0.45\textwidth]{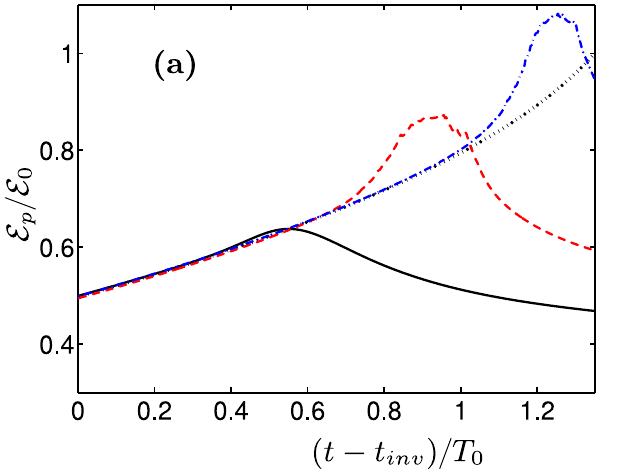}
    \includegraphics[width=0.44\textwidth]{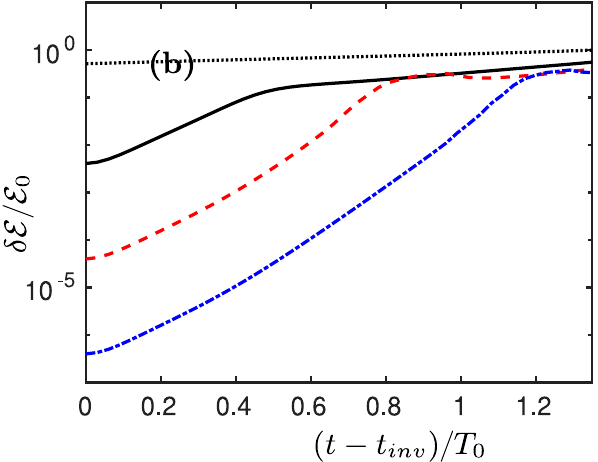}

    \includegraphics[width=0.46\textwidth]{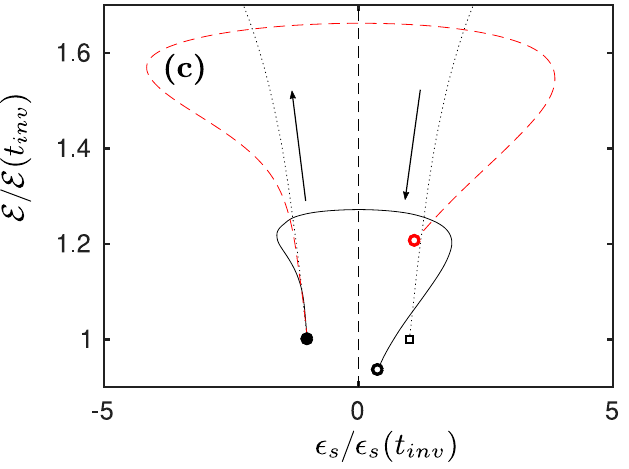}
    \includegraphics[width=0.44\textwidth]{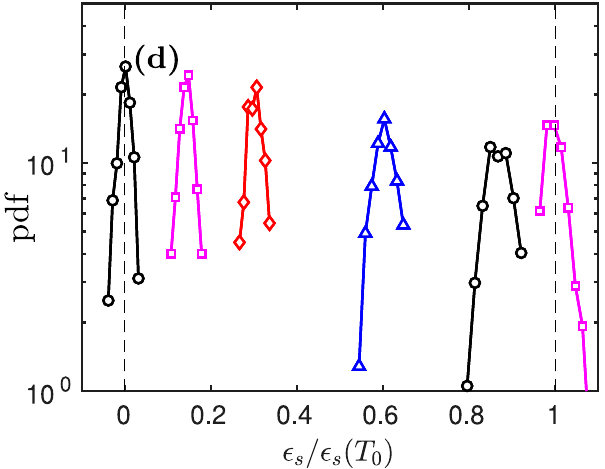}

      \caption{(a) Evolution of the energy of the perturbed fields $\mathcal{E}_p$ and (b) of $\delta \mathcal{E}=1/2\langle \delta{\boldsymbol{u}}^2\rangle$, normalised with $\mathcal{E}_0$, for: \solid,  \refc{$\mu=0.1$}; {\color{red} \dashed}, $\mu=0.01$; {\color{blue} \dotdashed}, $\mu=0.001$; \dotted, energy of the unperturbed trajectory, $\mathcal{E}(t)/\mathcal{E}_0$.
%       ; \solidcircle, $\mu=0.001$ in the direct evolution. The rate of growth of the perturbations are represented by the solid lines, where $\upsilon^+T_0\sim3.4$ and $\upsilon^-T_0\sim13.6$. 
      (c) Perturbed trajectories for $\mu=0.1$ and $\mu=0.01$, represented in the energy-dissipation space.
      Symbols as in (a). The solid circle represents the beginning of the inverse trajectory at $t_{inv}$, the empty circles the end of the perturbed trajectories at $2t_{inv}$, and the solid  square the end of the direct trajectory at time $t_{inv}$. The dotted lines and the arrows represent the direct (right) and the inverse (left) trajectories. (d) Evolution of the probability density function of $\epsilon_s$ as a function of time
      starting from random initial conditions at $t_{start}$. From left to right: $(t-t_{start})/T_0$=: $0.0$; $0.008$; $0.017$; $0.033$; $0.067$; $1.0$. $\epsilon_s$ normalised with the ensemble average at $T_0$.} 
    \label{fig:perturb}
  \end{center}
\end{figure}

% In $\S$\ref{sec:model}, we have shown that  the velocity field during the inverse evolution is similar to that in the direct evolution although with inverse sign, which is consistent with the time-symmetry of the equations governing the flow.
In the first place we confirm, by perturbing the inverse cascade, 
that in the neighbourhood of each inverse evolution there exist a
dense distribution of phase-space trajectories that also display inverse dynamics.
Each inverse trajectory is `unstable', in the sense that it is eventually destroyed when perturbed, 
but inverse dynamics are easily found considerably far from the original phase-space trajectories, even for distances of the order of the size of the accessible phase space, 
suggesting that inverse and direct trajectories lie in separated regions of phase space, rather than mostly being intertwined in the same neighbourhood.
\refc{Hereafter we refer to the phase-space of the Fourier coefficients, and consider, unless stated, experiments with the reversible SGS model}.

% Nonetheless, for short times the probability of observing states characterized by inverse dynamics increases and must be taken into consideration.
%which accounts for the irevesibility emerging from reversible dynamics.
%%%
%%Consequently, the dynamics of the inverse cascade are shown not to be unique of forward trajectories evolved backwards in time, but of a wider range of states in phase space.

In this experiment we generate a backwards initial condition by integrating the equations of motion until $t_{inv}$, changing the sign of the velocities $\boldsymbol{u}^\iota=-\boldsymbol{u}(t_{inv})$, and introducing a perturbation, $\delta \boldsymbol{u}$.
The perturbed flow field, $\boldsymbol{u}^p=\boldsymbol{u}^\iota+\delta \boldsymbol{u}$, is evolved and compared to the unperturbed trajectory, $\boldsymbol{u}^\iota(t)$.
We choose the initial perturbation field in Fourier space as,
\begin{equation}
\delta \widehat{u}_i({\boldsymbol{k}})= \widehat{u}^\iota_i({\boldsymbol{k}})\cdot \mu \exp(\ii \phi),
\label{eq:perturb}
\end{equation}
where $\phi$ is a random angle with uniform distribution between $0$ and $2\pi$ and $\mu$ is a real parameter that sets the initial energy of the perturbation.
The same $\phi$ is used for the three velocity components, so that incompressibility also holds for $\boldsymbol{u}^p$.
Figure \ref{fig:perturb}(a,b) shows the evolution of the mean kinetic energy of the perturbed field, $\mathcal{E}_p=\tfrac{1}{2}\langle {(\boldsymbol{u}^p)}^2 \rangle$, and of the perturbation field, $\delta \mathcal{E}=\tfrac{1}{2}\langle \delta\boldsymbol{u}^2\rangle$, for  $\mu=0.1$, $0.01$ and $0.001$. The energy of the perturbed fields increases for a considerable time in the three cases, demonstrating the presence of inverse energy transfer and negative mean eddy-viscosity, $C_s<0$. Even under the strongest initial perturbation,
$\mu=0.1$, the inverse cascade is sustained for approximately $0.5T_0$. 
\refb{Note that $\delta \mathcal E$ grows exponentially at a constant rate for some time under the linearised dynamics of the system,
indicating that perturbations tend to the most unstable linear perturbation \citep{benettin1980lyapunov}, which is independent of the form of the initial perturbation.}

As shown in figure \ref{fig:perturb}(b), by the time the inverse cascade is destroyed and the energy of the perturbed fields has evolved to a maximum and $C_s=0$, the energy of the perturbation is comparable to the total energy in all cases, $\delta \mathcal{E}\sim\mathcal{E}$,
which indicates that the states that separate inverse and direct dynamics are considerably far in phase space from the unperturbed trajectories, \refb{at least when the initial perturbations are small}. If we estimate distances as the square-root of the energy ($L_2$ norm of the velocity field), the maximum of $\mathcal{E}_p$ is located approximately at $\delta \mathcal{E}^{1/2}\simeq0.4\mathcal{E}_0^{1/2}$ from the unperturbed phase-space trajectory. In figure \ref{fig:perturb}(c) we represent the evolution of the perturbed trajectories in the dissipation-energy space. 
This plot intuitively shows that a typical direct trajectory is located far apart in phase space from inverse trajectories, and that perturbed inverse trajectories must first cross the set of states with zero dissipation, $C_s=0$, before developing a direct cascade.

In figure \ref{fig:perturb}(d) we show the evolution of the probability density function of the sub-grid energy transfer, $\epsilon_s$, in the direct evolution of the ensemble, starting at $t_{start}$, when the initial condition is a random field with a prescribed energy spectrum.
The initial probability distribution of $\epsilon_s$ is symmetric and has zero mean, indicating that the probabilities of direct and inverse evolutions are similar. As the system evolves in time and develops turbulence structure,
the probability of finding a negative value of $\epsilon_s$ decreases drastically.
At $t-t_{start}=0.008T_0$, we do not observe any negative values of $\epsilon_s$ in the ensemble. 
The standard deviation of $\epsilon_s$ is small compared to the mean at $t-t_{start}=T_0$, which indicates the negligible probability of observing negative values of 
$\epsilon_s$ after a time of the order of an eddy-turnover time.

Although decaying turbulent flows have no attractor in the strict sense, i.e. a set of phase-space points to which the system is attracted and
where it remains during its evolution, we use the term here to denote the set of phase-space trajectories with a fully turbulent structure and a direct energy cascade. 
As we have shown, this set of states is in fact `attracting', but the flow does not remain indefinitely turbulent due to its decaying nature. 
Let us note that it is possible to construct a reversible turbulent system with a proper turbulent attractor by including a reversible forcing in the large scales,
for instance a linear forcing that keeps the total energy of the system constant in time.

Inverse evolutions sustained in time are almost only accessible from forward-evolved flow fields with changed sign. These trajectories lie in the antiattractor, which is constructed by applying the transformation
$\boldsymbol u\rightarrow -\boldsymbol u$ to the turbulent attractor.
% antiattractor, which is composed of direct trajectories with the turbulent attractor
% which might seem to limit our analysis to very special trajectories.
However, we have shown in this analysis that inverse trajectories exist in a larger set of states outside the antiattractor, which can be escaped by perturbing the reversed flow fields.
%The destruction of the inverse cascade under perturbations, and the small standard deviation of $C_s$ in the turbulent attractor, reflect the negligible probability of inverse evolutions, and indicate that they can only be accessed temporarily.
\refc{The destruction of the inverse cascade under perturbations reflects the negligible probability of inverse evolutions, and indicates that they can only be accessed temporarily. 
Moreover, the small standard deviation of the sub-grid energy transfer with respect to the mean in the turbulent attractor
reflects the negligible probability that the system spontaneously escapes the attractor and develops and inverse energy cascade.}

% Although $C_s$ is not  proportional to the energy fluxes at all inertial scales, 
% (\ref{equ:fluxes}) shows that it is proportional to the energy fluxes at the grid filter, $\tau_{ij}S_{ij}=C_s\Delta^2|S|^3$.
% Since this identity holds at both the test and cut-off filter, the sign of $C_s$ reflects the direction of the cascade at both scales.
% Although inverse cascades are conceivable far from the test and cut-off filter, we have shown in $\S$\ref{sec:nomod} that the survival of the inverse cascade requires an energy supply in the small scales, which necessarily implies that $C_s<0$. 
% Thus the low probability of negative $C_s$ in the turbulent attractor reflects the low probability of sustained inverse energy cascades.}

The reversible turbulent system under study provides access to trajectories outside the turbulence attractor, which contains almost exclusively direct cascades.
By studying these inverse trajectories, we identify the physical mechanisms that lead to
the prevalence of direct over inverse cascades.

% lead trajectories to
% a turbulent attractor composed almost exclusively of  direct cascades.
% over inverse cascades in the turbulent attractor.
% that contains almost exclusively trajectories with an average direct energy cascade. 
% Access this trajectories

% \begin{figure}
% \centering
%     \psfrag{A}{\hspace{0pt}$\mathcal{E}/\langle\mathcal{E}\rangle_{erg}$}
%     \psfrag{B}{\hspace{0pt}$\epsilon_s/\langle\epsilon_s\rangle_{erg}$}
%     \includegraphics[width=0.4\textwidth]{attractor.pdf}%
%     \caption{(red) Phase-space trajectory of a long run 
%          of a microscopically reversible turbulent flow 
%          for $233T_{eto}$, where $T_{eto}$ is the eddy turn-over time.
%          (blue) Anti-trajectory obtained
%          by changing the sign of the dissipation. This trajectory can
%          be ideally obtained by changing the sign of a turbulent flow field within the attractor.
%          Here $E$ is the volume-averaged 
%          kinetic energy, $\epsilon$ is the volume-averaged dissipation and 
%          $\langle \cdot \rangle$ denotes ergodic phase-space average.
%          The shaded area represent states outside the turbulent attractor with average direct energy 
%          transfer, where we expect to find strange cascades.}
%     \label{fig:fluc}
% \end{figure}

\subsection{The structure of local energy fluxes in physical space}
\label{sec:spatio}
\begin{figure}
% \vspace{10pt}
\centering
    \includegraphics[width=0.48\textwidth]{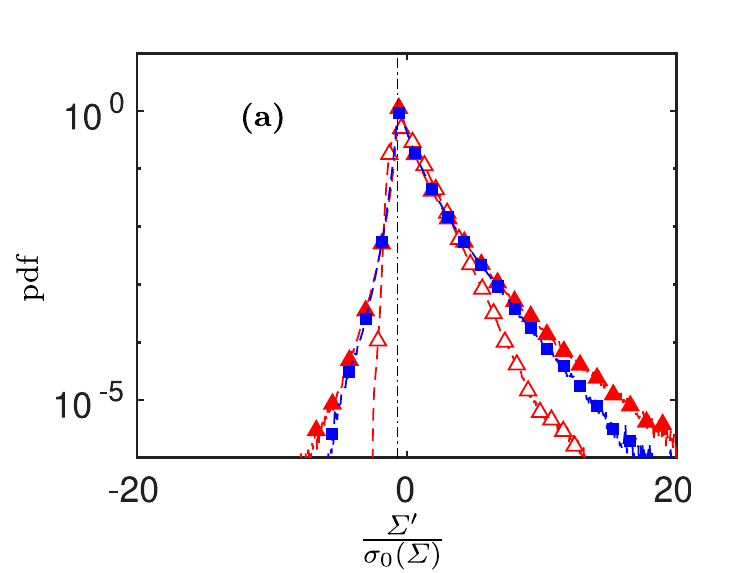}
    \includegraphics[width=0.45\textwidth]{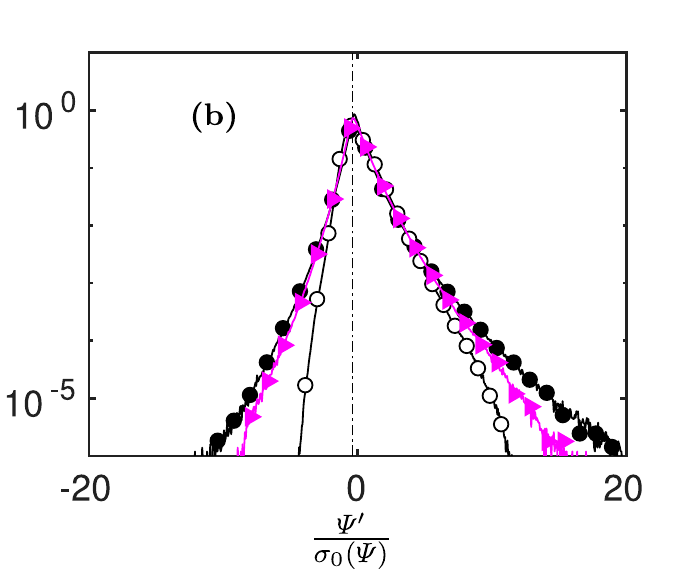}

          \caption{(a,b) Probability density function of local markers of the energy cascade without volume averaging: (a) $\Sigma$; (b) $\Psi$.
          %%%%%
          Symbols in correspond to:
%           \eta=$ {\protect\linesquare{red}}, 19;
%         \protect\linescircle{blue}, 38; \protect\linestriad{black}, 75; \protect\linestriad{magenta}
          {\protect\dashtriau{red}}, $\Sigma(5\Delta_{g})$;
          {\protect\dashquare{blue}}, $\Sigma(10\Delta_{g})$;
          {\protect\linestriar{magenta}}, $\Psi(5\Delta_{g})$;
          {\protect\linescircle{black}}, $\Psi(10\Delta_{g})$.
          The dashed-dotted lines represent $-\langle \psi \rangle/\sigma_V$ for quantities calculated at $5\Delta_{g}$, and the open symbols represent the probability density function of $\langle\Sigma(5\Delta_{g})\rangle_V$ and $\langle\Psi(5\Delta_{g})\rangle_V$ for $V=(16\Delta_{g})^3$.
          %%%
        %%%%%
%           and interscale auto-correlation coefficient as a function of the averaging volume:
%           {\protect\linestriau{black}}, $\mathcal{R}_{\langle\Sigma\rangle_V}(5\Delta_{g},10\Delta_{g})$; 
%           {\protect\linestriar{magenta}}, $\mathcal{R}_{\langle\Psi\rangle_V}(5\Delta_{g},10\Delta_{g})$;
%           {\protect\dashtriau{black}}, $\mathcal{R}^-_{\langle\Sigma\rangle_V}(5\Delta_{g},10\Delta_{g})$; 
%           {\protect\dashtriar{magenta}}, $\mathcal{R}^-_{\langle\Psi\rangle_V}(5\Delta_{g},10\Delta_{g})$.
%           and the dashed-dotted line represents $\log \mathcal P \sim 2{\langle \psi \rangle^2}/{\sigma_V^2}$. 
%           Error bars in (f) are calculated from the average standard deviation of $\log\mathcal{P}$ when partitioning the data-set in 4 subsets.
%           and the dashed-dotted line represents $\log \mathcal P \sim 3{\langle \psi \rangle^2}/{\sigma_V^2}$. 
%   %        
% % % % % %           Relation of $\alpha$ with $V/\Delta^3$ and $t/T_\Delta$ for \solid, $\Delta=\overline \Delta$; \dashed,$\Delta=2\overline \Delta$. (a) Dependence of $\alpha$ on $V/\Delta^3$ for $t/T_\Delta$ {\bf\Large\circle}, $0.33$; {\color{red}\solidtrian}, $0.66$; {\color{blue}\Large\solidrtrian2}, $1.05$; 
% %         {\color{magenta} \solidsquar}, $1.34$; {\Large\color{cyan} \soliddtrian1}, $1.93$.(b) Dependence of $\alpha$ on $t/T_\Delta$ for $V/\Delta^3$ {\bf\Large\circle}, $61$; {\color{red}\solidtrian}, $120$; {\color{blue}\Large\solidrtrian2}, $210$;{\color{magenta}\solidsquar}, $330$; {\color{cyan}\Large\soliddtrian1}, $495$;
        } 
    \label{fig:sigma_fig_1}
\end{figure}

Up to now we have only dealt with the mean transfer of energy over the complete domain. 
In this section we characterise the energy cascade as a local process in physical space.
We study here two markers of local energy transfer in physical space, 
$\Sigma(\boldsymbol x,t;\Delta)$ and $\Psi(\boldsymbol x,t;\Delta)$, previously defined in (\ref{eq:sigma}), calculated at filter scales $\Delta=5\Delta_{g}$ and $\Delta=10\Delta_{g}$.
Figures \ref{fig:sigma_fig_1}(a,b) show the probability distribution of these quantities in the direct evolution. 
We observe wide tails in the probability distribution of the two quantities, 
and skewness towards positive events, which is more pronounced for $\Sigma$ than for $\Psi$.
% As explained in These differences stem from the definition of both terms,
% which differ by the divergence of a spatial flux.
% Despite these differences, we will show that both quantities are robust indicators of the local direction 
Energy fluxes change sign under $\boldsymbol{u}\rightarrow-\boldsymbol u$, and, as shown in figure  \ref{fig:sigma_fig_1}, their odd-order moments are in general non-zero and have a definite sign in turbulent flows. This sign denotes statistical irreversibility, i.e, the privileged temporal direction of the system in an out-of-equilibrium evolution. The spatial average, which marks the direction of the cascade, is an important example, $\langle\Sigma\rangle=\langle\Psi\rangle>0$.

% reflecting the direction of the system in time in an out-of-equilibrium evolution.

To characterise the non-local spatial properties of the local energy fluxes,
we define the correlation coefficient between two scalar fields, $\psi$ and $\zeta$, as
\begin{equation}
\mathcal{S}_{\psi,\zeta}(\Delta \boldsymbol x)=\frac{\langle \psi'(\boldsymbol x + \Delta \boldsymbol x) \zeta'(\boldsymbol x)\rangle}{\sqrt{\langle \psi'^2\rangle \langle\zeta'^2\rangle}},
\label{corr}
\end{equation}
% \begin{equation}
% \mathcal{R}(\Delta x;\psi)=\frac{\langle \psi(\boldsymbol x + \Delta \boldsymbol x)\psi(\boldsymbol x)\rangle-\langle\psi(\boldsymbol x)\rangle^2}{\langle\psi(\boldsymbol x)^2\rangle-\langle\psi(\boldsymbol x)\rangle^2},
% \end{equation}
%%
% where $\psi$ and $\zeta$ stand as $\Sigma$ or $\Psi$.
where we subtract the spatial average from quantities marked with primes.  
Due to isotropy, $\mathcal{S}$ only depends on the spatial distance, $\Delta x=|\Delta \boldsymbol x|$. We also define the auto-correlation coefficient $\mathcal{S}_{\psi,\psi}(\Delta x)$ and the auto-correlation length as
\begin{equation}
\ell_\psi=\int^\infty_0 \mathcal{S}_{\psi,\psi}(\xi)\dd \xi,
\end{equation}
which measures the typical length over which $\psi$ decorrelates, and is related to the typical size of events in $\psi$.
The correlation lengths  of $\Sigma$ and $\Psi$ are proportional to the filter size $\Delta$, $\ell_\Sigma=0.47 \Delta$ and $\ell_\Psi=0.51 \Delta$ .

\begin{figure}
\vspace{10pt}
\centering
     \includegraphics[width=0.45\textwidth]{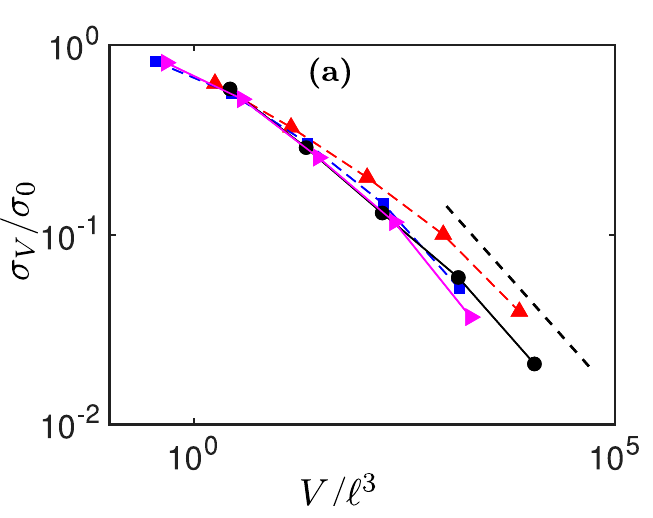}
     \includegraphics[width=0.46\textwidth]{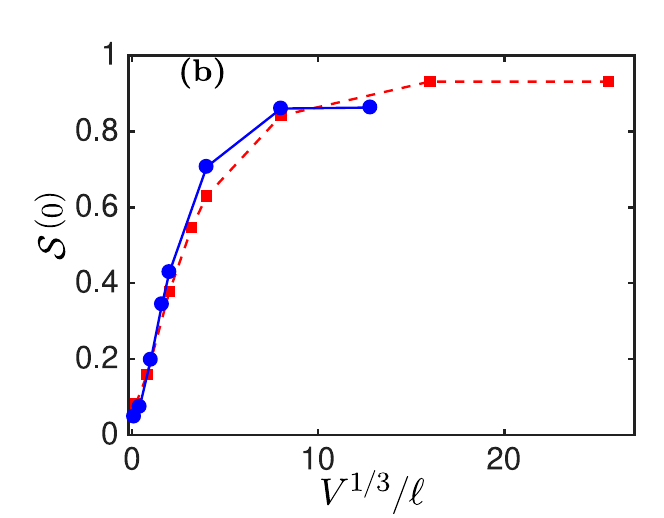}

     \includegraphics[width=0.45\textwidth]{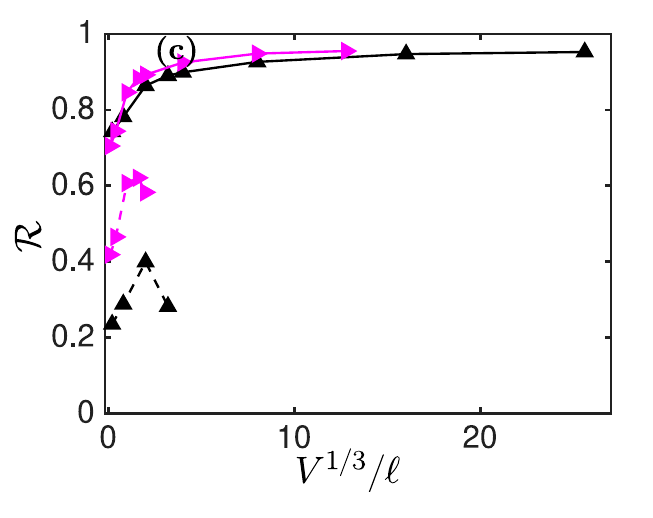}
     \includegraphics[width=0.46\textwidth]{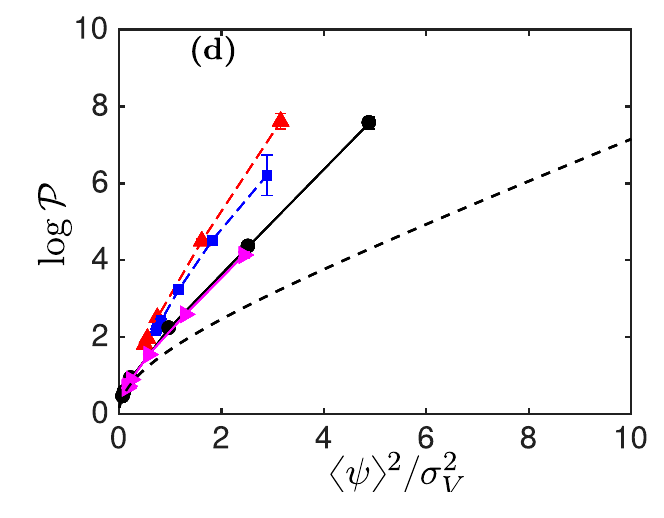}

     \includegraphics[width=0.45\textwidth]{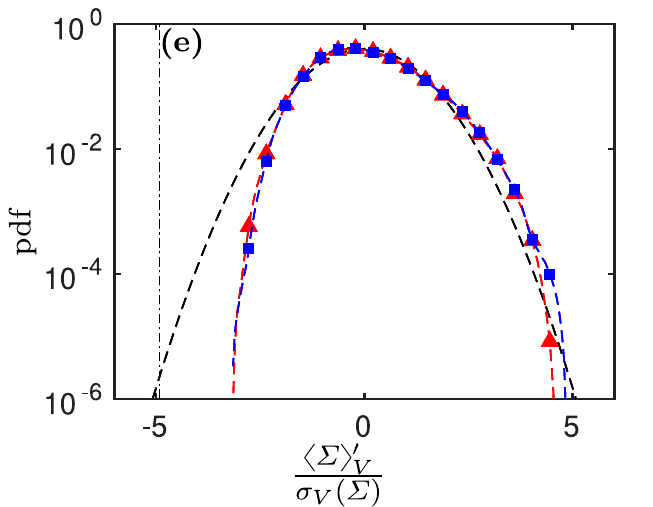}
     \includegraphics[width=0.46\textwidth]{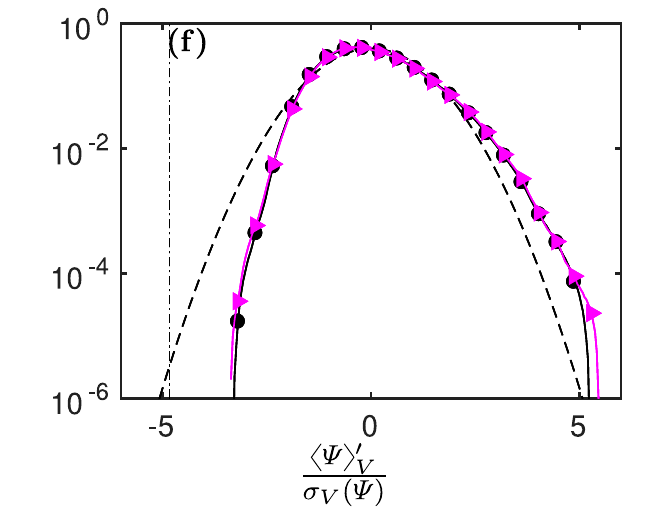}

          \caption{(a) $\sigma_V$ as a function of $V/\ell^3$, where the dashed line represents $\sigma_V\sim \mathcal{N}^{-1/2}$. Symbols correspond to:
%           \eta=$ {\protect\linesquare{red}}, 19;
%         \protect\linescircle{blue}, 38; \protect\linestriad{black}, 75; \protect\linestriad{magenta}
          {\protect\dashtriau{red}}, $\Sigma(5\Delta_{g})$;
          {\protect\dashquare{blue}}, $\Sigma(10\Delta_{g})$;
          {\protect\linestriar{magenta}}, $\Psi(5\Delta_{g})$;
          {\protect\linescircle{black}}, $\Psi(10\Delta_{g})$.
          (b) Correlation coefficient between $\Sigma$ and $\Psi$ as a function of the averaging volume: 
          {\protect\dashquare{red}}, $\mathcal{S}_{\langle\Sigma\rangle_V,\langle\Psi\rangle_V}(0)$ at  $5\Delta_{g}$;  
          {\protect\linescircle{blue}}, $\mathcal{S}_{\langle\Sigma\rangle_V,\langle\Psi\rangle_V}(0)$
          at $10\Delta_{g}$.
%           and the dashed-dotted line represents $\log \mathcal P \sim 2{\langle \psi \rangle^2}/{\sigma_V^2}$. 
          Error bars are calculated from the average standard deviation of $\log\mathcal{P}$ when partitioning the complete data-set in 4 subsets.
          (c) Interscale auto-correlation coefficient as a function of the averaging volume:
          {\protect\linestriau{black}}, $\mathcal{R}_{\langle\Sigma\rangle_V}(5\Delta_{g},10\Delta_{g})$; 
          {\protect\linestriar{magenta}}, $\mathcal{R}_{\langle\Psi\rangle_V}(5\Delta_{g},10\Delta_{g})$;
          {\protect\dashtriau{black}}, $\mathcal{R}^-_{\langle\Sigma\rangle_V}(5\Delta_{g},10\Delta_{g})$; 
          {\protect\dashtriar{magenta}}, $\mathcal{R}^-_{\langle\Psi\rangle_V}(5\Delta_{g},10\Delta_{g})$.
          (d) Asymmetry function, $\mathcal P$, as a function of $\langle \psi \rangle^2/\sigma^2_V$. \refb{Symbols as in (a)}. 
          %%%%%
          The dashed line is the exact solution of $\mathcal{P}$ for a Gaussian, which approaches $\log \mathcal P=1/2{\langle \psi \rangle^2}/{\sigma_V^2}$ for large ${\langle \psi \rangle^2}/{\sigma_V^2}$.
          (e,f) Probability density function of local markers of the energy cascade averaged at scale $V^{1/3}=1.2L_0=32\Delta_g$: (e) $\Sigma$; (f) $\Psi$. 
          The dashed-dotted lines represent $-\langle \psi \rangle/\sigma_V$ for quantities calculated at $5\Delta_{g}$. Symbols as in (a).
%           and the dashed-dotted line represents $\log \mathcal P \sim 3{\langle \psi \rangle^2}/{\sigma_V^2}$. 
% % % % % %           Relation of $\alpha$ with $V/\Delta^3$ and $t/T_\Delta$ for \solid, $\Delta=\overline \Delta$; \dashed,$\Delta=2\overline \Delta$. (a) Dependence of $\alpha$ on $V/\Delta^3$ for $t/T_\Delta$ {\bf\Large\circle}, $0.33$; {\color{red}\solidtrian}, $0.66$; {\color{blue}\Large\solidrtrian2}, $1.05$; 
% %         {\color{magenta} \solidsquar}, $1.34$; {\Large\color{cyan} \soliddtrian1}, $1.93$.(b) Dependence of $\alpha$ on $t/T_\Delta$ for $V/\Delta^3$ {\bf\Large\circle}, $61$; {\color{red}\solidtrian}, $120$; {\color{blue}\Large\solidrtrian2}, $210$;{\color{magenta}\solidsquar}, $330$; {\color{cyan}\Large\soliddtrian1}, $495$;
        } 
    \label{fig:sigma_fig_2}
\end{figure}

% .$\Psi$ has a finer spatial structure than $\Sigma$, probably due to the fact that $\Psi$ is calculated from a product and a division of intermittent quantities.
We use $\langle \Sigma \rangle_V$ and $\langle \Psi \rangle_V$ to  study the spatial structure of energy fluxes, where $\langle \cdot \rangle_V$ represents the volume-averaging operation over a sphere of volume $V$. The standard deviation of the probability distribution of the volume-averaged fields,
%%%
\begin{equation}
\sigma_V=\sqrt{\langle\big(\langle\psi\rangle_V-\langle\psi\rangle\big)^2\rangle},
\end{equation}
%%%
is shown in figure \ref{fig:sigma_fig_2}(a) as a function of the averaging volume, where $\sigma_0=(\langle \psi^2\rangle - \langle \psi \rangle^2)^{1/2}$ is the standard deviation of the test field without volume averaging.
When the averaging volume is sufficiently large, $V/\ell^3\gtrsim10^3$, the standard deviation
becomes inversely proportional to the square-root of $\mathcal{N}=V/\ell^3$ in all cases,
where $\mathcal{N}$ is a measure of the number of independent energy transfer events within the averaging volume. These results suggest statistical independence of the events within the averaging volumes and, some degree of space locality in the energy cascade.
% that as we consider averaging volumes such that $\mathcal{N}\gg1$, the prob

% Despite the differences between $\Sigma$ and $\Psi$, we now show that both markers are similar
% when locally averaged, indicating that the energy cascade can be robustly characterised 
% regardless of the particular definition of the energy fluxes. 
% We define the correlation coefficient between two fields as 
% %%
% \begin{equation}
% \mathcal{S}(\psi,\zeta)=\frac{\langle(\psi(\boldsymbol x) - \langle \psi(\boldsymbol x) \rangle) (\zeta(\boldsymbol x) - \langle \zeta(\boldsymbol x)\rangle)\rangle}{\sqrt{\langle (\psi(\boldsymbol x) - \langle \psi \rangle)^2\rangle \langle(\zeta(\boldsymbol x) - \langle \zeta(\boldsymbol x) \rangle)^2\rangle}}
% \label{corr}
% \end{equation}
% %%
% where the average is performed over all the domain.
Despite this locality, $\Sigma$ and $\Psi$ are very different pointwise, which might suggest
that local energy fluxes are not uniquely defined, and that only global averages are robust with respect to the particular definition of fluxes,
$\langle\Sigma\rangle=\langle\Psi\rangle$. In fact, the correlation between $\Sigma$ and $\Psi$ for $\Delta x=0$ is low, $\mathcal{S}_{\Sigma,\Psi}(0)\sim0.05$, at both filter scales, $5\Delta_{g}$ and $10\Delta_{g}$.
However, when these quantities are averaged, the correlation coefficient increases rapidly with the averaging volume. In figure \ref{fig:sigma_fig_2}(b), we show the dependence of $\mathcal{S}_{\langle\Sigma\rangle_V, \langle \Psi\rangle_V}(0)$ on $V$. For averaging volumes $V\sim(4\ell)^3$, the correlation increases to approximately $0.7$ for both filter widths, indicating that $\Sigma$ and $\Psi$ are similar when averaged over volumes 
of the order of their cubed correlation length.
% From now on, we consider only correlations without spatial increments $\Delta x=0$.
% % These results indicate that the $\Sigma$ and $\Psi$ are robust markers of the direction and the intensity of the local energy fluxes on a local average sense. 

% We also define the conditional correlation,
% $\mathcal{S}_{\psi,\zeta|A}$, which is similar to (\ref{corr}), except that averages are taken only in points fulfilling a prescribed condition $A$.

\begin{table}  
\begin{center}
\def~{\hphantom{0}}
\begin{tabular}{c c  c  c}
                                             & `All'    & `$+$'    & `$-$'   \\ 
 $\mathcal{R}_\Sigma(5\Delta_g,10\Delta_g)$  & $0.74$ & $0.73$ & $0.2$ \\ 
 $\mathcal{R}_\Psi(5\Delta_g,10\Delta_g)$    & $0.70$ & $0.60$ & $0.4$ \\ 
\end{tabular}
\hspace{20pt}
\begin{tabular}{c c  c  c}
                                             & `All'    & `$+$'    &  `$-$'   \\ 
 $\mathcal{R}_\Sigma(5\Delta_g,20\Delta_g)$  & $0.33$ & $0.34$ &  $-0.01$ \\ 
 $\mathcal{R}_\Psi(5\Delta_g,20\Delta_g)$    & $0.24$ & $0.29$ &  $-0.09$ \\ 
\end{tabular}
\caption{Auto-correlation coefficient among scales of $\Sigma$ and $\Psi$, evaluated for: `All', the complete field; `$+$', conditioned to positive energy transfer events at $5\Delta_g$; `$-$', conditioned to negative energy transfer events at $5\Delta_g$.}
\label{tab:kd2}
\end{center}
\end{table}

\subsection{The structure of local energy fluxes in scale space}
\label{sec:spatio2}

We extend this analysis to scale space by considering the correlation of the energy
fluxes at different scales, defined as
\begin{equation}
\mathcal{R}_\psi(\Delta_1,\Delta_2)=\frac{\langle \psi'(\boldsymbol x;\Delta_1)\psi'(\boldsymbol x;\Delta_2)\rangle}{\sqrt{\langle (\psi'(\Delta_1))^2\rangle \langle(\psi'(\Delta_2))^2\rangle}},
\label{corr}
\end{equation}
and calculate the same quantity conditioning the averages to positive or negative energy transfer events at scale $\Delta_1$. We denote the conditional correlations by
$\mathcal{R}^+$ when conditioning to $\psi(\Delta_1)>0$, and $\mathcal{R}^-$ when conditioning
to $\psi(\Delta_1)<0$. These interscale auto-correlation coefficients are presented in table \ref{tab:kd2}. We find correlation values of approximately $0.7$ for scale increments of $2$ in both quantities. Table \ref{tab:kd2} also includes results for $\Delta_2=20\Delta_g$. It shows that all fluxes substantially decorrelate for $\Delta_2/\Delta_1=4$ \citep{cardesa2017turbulent}, and that this is especially true for the backscatter.  

% \begin{equation}
% \begin{aligned}
% \mathcal{R}^+_\psi(\Delta_1,\Delta_2)&=\mathcal{S}_{\psi(\Delta_1),\psi(\Delta_2)}|\psi(\Delta_1)>0,\\
% \mathcal{R}^-_\psi(\Delta_1,\Delta_2)&=\mathcal{S}_{\psi(\Delta_1),\psi(\Delta_2)}|\psi(\Delta_1)<0,
% \end{aligned}
% \end{equation}
% which shows that $\Sigma$ and $\Psi$ are highly correlated in scale. 
% When we increase the scale gap we find lower correlations, 
% $\mathcal{R}_\Sigma(5\Delta_{g}, 20\Delta_{g})=0.3$ and
% $\mathcal{R}_\Psi(5\Delta_{g}, 20\Delta_{g})=0.23$.
% When we condition the correlation coefficient to direct energy transfer,
% we also find strong correlation, $\mathcal{R}^+_\Sigma(5\Delta_{g},10\Delta_{g})=0.73$ and $\mathcal{R}^+_\Psi(5\Delta_{g},10\Delta_{g})=0.66$. However, when we condition to inverse energy transfer we obtain weaker correlations, $\mathcal{R}^-_\Sigma(5\Delta_{g},10\Delta_{g})=0.2$ and $\mathcal{R}^-_\Psi(5\Delta_{g},10\Delta_{g})=0.4$. 

Figure \ref{fig:sigma_fig_2}(c) shows the dependence of $\mathcal{R}_{\langle\Sigma\rangle_V}(5\Delta_{g}, 10\Delta_{g})$ and $\mathcal{R}_{\langle\Psi\rangle_V}(5\Delta_{g},10\Delta_{g})$ on the averaging volume $V$. 
The interscale auto-correlation increases with the averaging volume when we consider all direct events (not shown).
On the other hand, the correlations conditioned to backscatter increase up to volumes of the order of $V^{1/3}\approx 2\ell$, and then decrease. 
Beyond $V^{1/3}\approx4\ell$, the number of averaged inverse energy transfer events is not enough to compute reliable correlations.

% We also observe that, while the interscale autocorrelation of the direct energy

% In conclusion, we have shown that the energy cascade can be properly characterised in physical space: it is approximately local and robust to the definition of the energy fluxes, on an average sense, over volumes of the order of the cubed filter width.
% % We have also analysed the structure of energy fluxes in scale space.
% Local energy fluxes at scales separated by a factor of $2$ are correlated, 
% while, in agreement with the reported scale-space locality of the cascade \citep{eyink2005locality,eyink2009localness,domaradzki2009locality,cardesa2017turbulent},
% the correlation between energy fluxes at scales separated by a factor of $4$ is weaker. 
% When conditioning correlations to the local direction of the cascade, 
% we find that the interscale correlation is strong for direct energy fluxes and  weak for energy backscatter. While direct energy fluxes result from the dynamic interaction
% between different scales, energy backscatter is a highly scale-local phenomena.

% while direct energy transfer is strongly auto-correlated at scales separated by a factor of $2$.

% This is probably due to the difference in the skewness of the probability distribution,
% which is higher for $\Sigma$ than for $C_s$. 
% .i.e we find more independent events per volume.
% ability of observing that volume $V$ develop an inverse cascade can be approximated asymptotically from the statistical distribution of single energy transfer events.

\subsection{Physical-space estimates of the probability of inverse cascades}

We have shown in $\S$\ref{sec:per} that the probability of spontaneously observing an inverse cascade over extended regions of a turbulent flow is negligible, making it impractical to quantify the probability of such evolutions by direct observation. 
% 
% % Despite the low probability of inverse cascades, we frequently observe local inverse energy transfer events over restricted regions of physical space \citep{piomelli1991subgrid,meneveau1991analysis,cerutti1998intermittency,ishihara2009study}.
However, as shown in $\S$\ref{sec:spatio}, we frequently observe local inverse energy transfer events over restricted regions of physical space. 
%  
% 
% is quantifiable
%%%
% These events are less likely and less intense than  direct energy transfer events, but their probability is quantifiable.
% % % and they have practical consequences, at least numerically.
% Here we present estimates on the probability of global inverse cascades by averaging these local events over large volumes, and taking the limit when the averaging volume becomes comparable to the flow domain.
%%%
We attempt to estimate the probability of observing an inverse cascade using
the integral asymmetry function, 
\begin{equation}
\mathcal{P}=\frac{P(\langle\psi\rangle_{V}>0)}{P(\langle\psi\rangle_{V}<0)},
\label{eq:asyi}
\end{equation}
which compares the probability of observing a direct to an inverse cascade in a volume $V$. 
This approach follows the methodology of the fluctuation relations \citep{evans2002fluctuation} and its local versions \citep{ayton2001local, michel2013local}.
% In the asymptotic limit in which $V\rightarrow L_0^3$ and $\Delta/L_0\ll1$, we approximate the probability of an inverse energy cascade over the complete domain at the inertial scale $\Delta$.

We evaluate the integral asymmetry function of $\Sigma$ and $\Psi$ for different averaging volumes, 
and display it as a function of $\langle \psi \rangle^2/\sigma_V^2$ in figure \ref{fig:sigma_fig_2}(d). 
%%
% For $\Psi$, we observe that, in the limit of $\mathcal{N}$ sufficiently large, $\log\mathcal{P}$ follows the trend of a Gaussian distribution with non-zero mean, $\langle \psi \rangle\neq0$, and standard deviation $\sigma_V$, 
% \begin{equation}
% \log \mathcal P=\frac{1}{2}\frac{\langle \psi \rangle^2}{\sigma_V^2}.
% \label{eq:gaussp}
% \end{equation}
% For $\langle \Sigma \rangle_V$ and ,
%%
%%
% The probability of average inverse energy transfer 
% decreases considerably with the averaging volume. 
The statistical independence of the energy transfer events reported in $\S$\ref{sec:spatio} suggests that, when the averaging volume is large enough,
the integral asymmetry function should behave as that of a Gaussian distribution with non-zero mean, which for large $\langle \psi \rangle^2/\sigma_V^2$ is 
\begin{equation}
\log \mathcal P\simeq2\langle \psi \rangle^2/\sigma_V^2.
\label{eq:prefactor}
\end{equation}
If we assume that the number of independent events in a volume $V$ is $\mathcal N\sim V/\ell^3$,
we can conclude that $\log P \sim \mathcal{N} \sim V/\ell^3$, 
and the probability of direct over inverse cascades increases exponentially with the number of 
independent energy transfer events considered.
The exponential dependence in (\ref{eq:prefactor}) is well satisfied in figure \ref{fig:sigma_fig_2}(d), but the prefactor is not, and $\log \mathcal P>2\langle \psi \rangle^2/\sigma_V^2$. 
Since the prefactor is a property of the Gaussian distributions that arise from the application of the central-limit theorem, 
this failure can be traced to the strong dependence of $\mathcal P$ on the negative tails of the
statistical distributions of $\Sigma$ and $\Psi$, which are not Gaussian.
The fast growth of $\log \mathcal P$ in figure \ref{fig:sigma_fig_2}(d) signals that the probability of inverse-transfer events
decreases with volume averaging faster than what could be expected for independent events,
in agreement with the evidence in figure \ref{fig:sigma_fig_2}(c) that inverse energy transfer is shallow in scale space. 
In figures \ref{fig:sigma_fig_1}(a,b), we show the probability density function of $\langle \Sigma(5\Delta_g) \rangle_V$ and $\langle \Sigma(5\Delta_g) \rangle_V$ for  $V=(16\Delta_g)^3$.
The negative tails decrease considerably with the averaging, and do not collapse with those of $\Sigma(10\Delta_g)$, and $\Psi(10\Delta_g)$, while the positive tails are less affected.
This asymmetry in the tails persists for large averaging volumes, as shown in figures \ref{fig:sigma_fig_2}(e,f), 
where we display the probability distribution of $\langle \Sigma \rangle_V$ and $\langle \Psi\rangle_V$ for a volume of the order of the integral scale of the flow, 
$V^{1/3}=1.2L_0=32\Delta_g$. Neither of the two probability distributions are Gaussian, specially in its negative tails.

These results point to essential differences between inverse and direct energy transfer events, which seem to
emerge from different dynamical processes. 
While direct energy fluxes are correlated with fluxes at larger scales, inverse energy fluxes are not, suggesting
that local backscatter, measured as $\Sigma<0$ or $\Psi<0$, does not necessarily imply a net inverse cascade of energy.
This is corroborated by the strong depletion of inverse energy transfer events with volume-averaging, which suggest that backscatter might
be a consequence of space-local conservative fluxes, which cancel out when averaged, rather than of interscale interactions.
Let us note that $\Sigma$ and $\Psi$ differ in fact by a spatial flux, and
that the intensity and probability of backscatter events are a specially noticeable difference between the two definitions.
These space-local fluxes seem to cause a higher probability of backscatter in $\Psi$ than in $\Sigma$,
suggesting that they are also, to some extent, the cause of backscatter in $\Sigma$. 
In view of these results, it is unlike that the probability of inverse cascades can be quantified from the probability of local backscatter.
Even if we could do so by analysing a massive database, and by devising a local definition of the energy transfer free of spatial fluxes,
our results suggests that the energy cascade is sufficiently out of equilibrium for these predictions to have little practical importance.

\section{The structure of the inverse and direct cascades}
\label{sec:qr}

% is resilient to disturbances from surrounding turbulence.
The velocity gradients constitute a convenient descriptor of the fundamental structure of turbulent flows and some of their statistics reflect the out-of-equilibrium nature of turbulence. 
Although they describe only the structure of the small scales,
the statistical distributions of the filtered velocity gradients are invariant across scales and reveal the self-similar structure of the inertial range \citep{borue1998local, luthi2007lagrangian, lozano2016multiscale, danish2018multiscale}. 
Some investigations suggest a connection between the velocity gradients in the inertial scales and the energy cascade \citep{borue1998local,meneveau1999conditional,goto2008physical}, and common SGS models rely on the assumption that energy transfer towards the unresolved scales can be reproduced using the velocity gradients of the resolved scales \citep{smagorinsky1963general, bardina1980improved,nicoud1999subgrid}.
% From a general perspective, the energy cascade is a process that generates velocity gradients at smaller scales until these gradients are strong enough to dissipate energy. 
% The analysis of this process is simplifiedd  the vorticity vector and the rate-of-strain tensor. 
The structure of the velocity gradients is compactly represented by the invariants of the velocity gradient tensor. Supported on the space-locality of the energy cascade reported in $\S$\ref{sec:spatio}, 
we will relate these invariants, which are strictly local in space, with local energy fluxes,
connecting the energy cascade in physical space to the local structure of the flow.

\subsection{Dynamics of the invariants of the velocity gradient tensor}
\label{sec:dyninv}
%

% This analysis reveals possible mechanisms for the energy cascade and allows us to established an entropic argument for the prevalence of direct energy transfer based on the dynamics of the flow topologies.
% This approach based on a local analysis of the flow topology is not new and references on the

% We expect the energy cascade to be an approximately local process in scale and physical space 
Let $\mathcal{A}_{ij}=\partial_{j} u_i$ be the velocity gradient tensor of an incompressible flow. The second and third invariants of $\mathcal{A}_{ij}$ are 
%
%and offer complete information on the geometry of flow.
%A comparison between the structure of the direct and inverse cascade can be established through these invariants. 
%From the velocity gradient tensor $A_{ij}=\partial_j u_i$ has two non-zero invariants $Q$ and $R$ defined as: 
%
\begin{align}
Q =&-\frac{1}{2}\mathcal{A}_{ij}\mathcal{A}_{ji}=\frac{1}{4}\omega_i\omega_i-\frac{1}{2}S_{ij}S_{ij},\label{eq:Q} \\
R =&-\frac{1}{3}\mathcal{A}_{ij}\mathcal{A}_{jk}\mathcal{A}_{ki}= -\frac{1}{3}S_{ij}S_{jk}S_{ki} - \frac{1}{4}\omega_iS_{ij}\omega_j, \label{eq:R}
\end{align}
where $\omega_i$ is the $i$-th component of the vorticity vector.
Considering $R$ and the discriminant $D=27/4R^2+Q^3$,
the flow can be classified in four different topological types:
positive $D$ %$Q^3>-27R^2/4$
corresponds to rotating topologies and 
negative $D$ to saddle-node topologies; 
negative $R$ accounts for topologies with a stretching principal direction, 
and positive $R$ for topologies with a compressing principal direction. 
These invariants are also related to the vorticity vector and the rate-of-strain tensor, such that $Q$ indicates the balance between enstrophy and strain, denoted by $|S|^2=2S_{ij}S_{ij}$ and $|\omega|^2=\omega_i\omega_i$, and $R$ represents the balance between vortex stretching and strain self-amplification, $\omega_iS_{ij}\omega_j$ and $S_{ij}S_{jk}S_{ki}$.
These terms appear in the evolution equations of $|S|^2$ and $|\omega|^2$,
\begin{align}
\frac{1}{4}D_t |S|^2     &=-S_{ij}S_{jk}S_{ki}-\frac{1}{4}\omega_iS_{ij}\omega_j-S_{ij}\partial_{ij}p,\label{equ:ss} \\
\frac{1}{2}D_t |\omega|^2&= \ \ \omega_iS_{ij}\omega_j,\label{equ:ome}
\end{align}
where $D_t=\partial_t+u_j\partial_j$ is the substantial derivative along a Lagrangian trajectory and we have considered the evolution of an inviscid flow for simplicity.
 
\begin{figure}
  \includegraphics[width=1.0\textwidth]{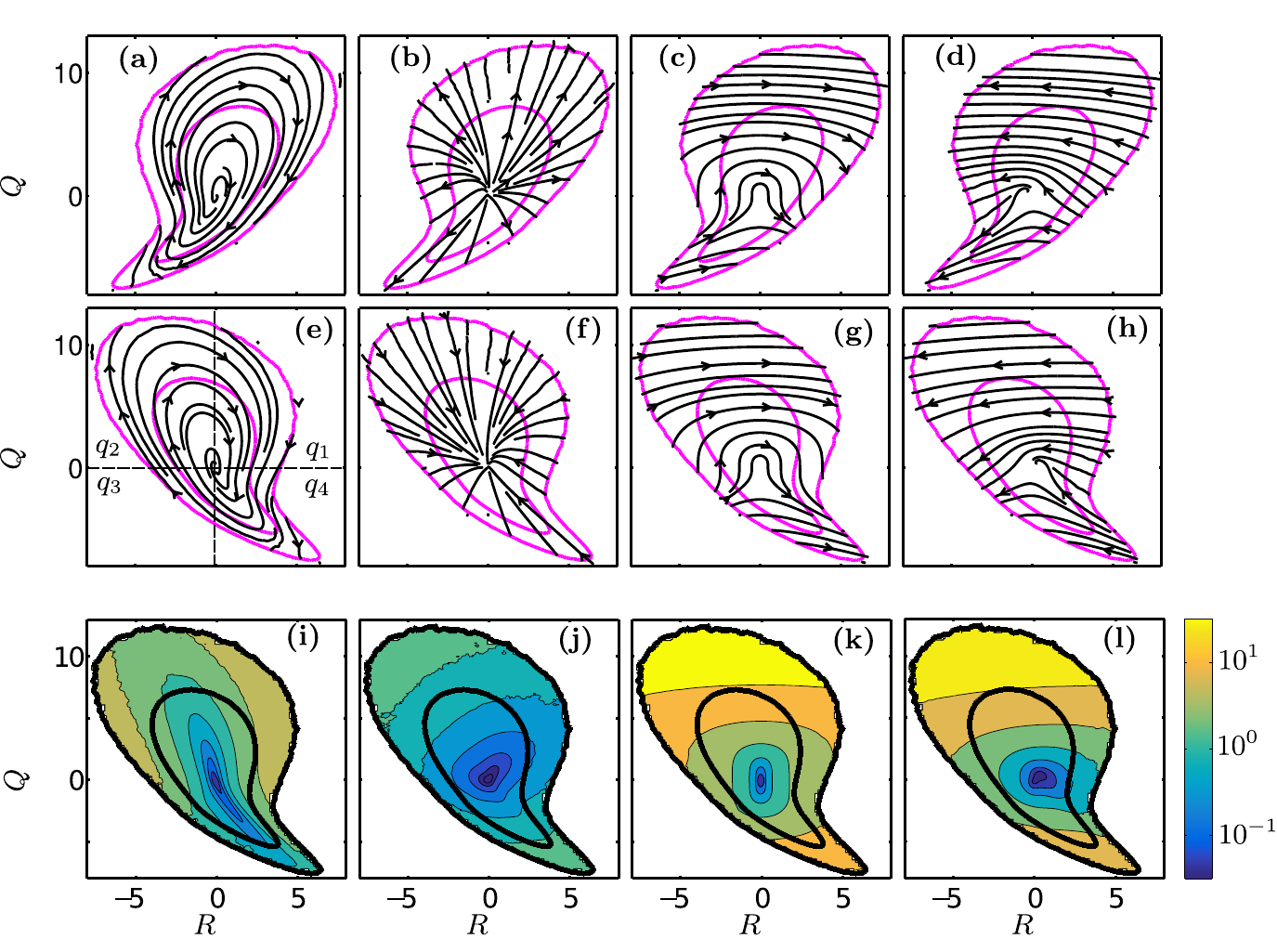}%
  %%%
  \caption{\protect\linearrow, conditional mean trajectories in the $Q$--$R$ plane for the inverse (a-d) and direct (e-h) evolutions due to different contributions. 
  (i-l) Modulus of the probability transport velocities $|\boldsymbol{\Phi}|$ in the $Q$--$R$ plane in the  direct evolution due to the different contributions.
%   Intensities in the inverse evolution are similar except for the transformation $R\rightarrow-R$.
  (a,e,i) All contributions $\boldsymbol{\Phi}_T$, 
  (b,f,j) SGS model $\boldsymbol{\Phi}_M$, 
  (c,g,k) restricted Euler $\boldsymbol{\Phi}_E$ and
  (d,h,l) non-local component of the pressure Hessian $\boldsymbol{\Phi}_P$.
  Contours of the probability density function of $Q$ and $R$ contain  0.9 and 0.96 of the total data.
  All quantities are normalised using $t_Q={\langle Q^2 \rangle}^{1/4}$.}
  \label{fig:lyap_QR_con}
\end{figure}

The statistical distribution of $Q$ and $R$ in turbulent flows has a typical teardrop shape, which is shown in figure \ref{fig:lyap_QR_con}(e). For ease of reference we have divided the $Q$-$R$ plane in four quadrants, where $q_1$ corresponds to $Q>0$ and $R>0$, $q_2$ to $Q>0$ and $R<0$, $q_3$ to $Q<0$ and $R<0$, and $q_4$ to $Q<0$ and $R>0$.
The teardrop shape is characterised by a lobe in $q_2$, where the enstrophy is dominant over the strain and the vortex stretching over the strain self-amplification,
and a tail in $q_4$, which is known as the \citet{vieillefosse1984internal} tail, and represents dominant strain self-amplification over vortex stretching in strain-dominated regions.
% Here we refer to strain production or strain self-amplification as $S_{ij}S_{jk}S_{ki}<0$ and to vortex stretching as $\omega_iS_{ij}\omega_j>0$.
% Let us note that the interpretation of $Q$ and $R$ in relation to $|S|$ and $|\omega|$ is not straightforward.
The high absolute values of $Q$ in the Vieillefosse tail and in the upper semiplane indicate the spatial segregation of $|S|$ and $|\omega|$.
Low absolute values of $Q$ do not in general imply low values of $|S|$ or $|\omega|$, but rather that $|S|\sim|\omega|$.
We actually find that $|S|\sim\langle |S| \rangle$ in regions where $Q\sim0$ and $R\sim0$.
% On the other hand high values of $Q$ denotes spatial segregation of $|S|$ and $|\omega|$.
% A similar analysis is necessary for $R$, which is even more difficult to interpret given that both addends can take positive and negative values.
For a detailed interpretation of the $Q$-$R$ plane refer to \citet{tsinober2000vortex}.
% We find that $Q$ and $R$ condense a significant amount of information in a compact representation and are useful quantities provided that we apply a rigorous interpretation. 

% \sqrt{S_{ij}S_{ij}}}$

The dynamics of the velocity gradients can be statistically represented by the probability transport velocities in the $Q$--$R$ plane,
which are obtained from the average Lagrangian evolution of $Q$ and $R$ \citep{ooi1999study}.
Taking spatial derivatives in the evolution equations of the velocity field described by (\ref{eq:1}) and considering (\ref{eq:Q}, \ref{eq:R}), the Lagrangian evolution of the invariants reads
\begin{align}
{D_tQ}=&-3R+\mathcal{A}_{ij}\mathcal{H}_{ji}, \\
{D_tR}=&\frac{2}{3}Q^2-\mathcal{A}_{ij}\mathcal{A}_{jk}\mathcal{H}_{ki},
\label{eq:dqdr}
\end{align}
where $\mathcal{H}_{ij}=\mathcal{H}^P_{ij}+\mathcal{H}^M_{ij}$ comprises the contribution of the non-local component of the pressure Hessian and the SGS model, $\mathcal{H}_{ij}^P={\partial_{ij} p -\frac{1}{3}(\partial_{kk}p)\delta_{ij}}$ and $\mathcal{H}_{ij}^M=\partial_j{M_i}$. 
Here, $M_i$  is the $i$-th component of the SGS model in the momentum equation,
and $\delta_{ij}$ is the Kronecker's delta.
The total probability transport velocity is defined as
\begin{equation}
\boldsymbol{\Phi}_T=\{\langle {D_tQ}\rangle_C t_Q^3, \langle {D_tR}\rangle_C t_Q^4\},
\end{equation}
where $\langle \cdot \rangle_C$ is the probability conditioned to $Q$ and $R$, and $t_Q=1/\langle Q^2 \rangle^{1/4}$
is a characteristic time extracted from the standard deviation of $Q$.
Let us note that the probability density flux in the $Q$-$R$ plane is $P(Q,R)\boldsymbol \Phi_T$, where $P(Q,R)$ is the joint probability density of $Q$ and $R$.
The probability transport velocities are integrated to yield conditional mean trajectories (CMTs) in the $Q$--$R$ plane \citep{ooi1999study}, which are shown also in figure \ref{fig:lyap_QR_con}(e).

% The CMTs develop an average rotating cycle characteristic of turbulent flows, which exposes causality between the different configurations of the flow in an average sense.
% The CMTs rotate clockwise: from vortex stretching to vortex compression in the upper semiplane, to the Vieillefosse tail in $q_4$, and again to vortex stretching in enstrophy-dominated regions.
% For closed steady states, CMTs rotate in a closed motion \citep{lozano2016multiscale}, while for closed decaying flows, they spiral inwards. 
% The CMTs are always found to rotate clockwise, from vortex stretching to vortex compression, to the Vieillefosse tail and again to vortex stretching and enstrophy dominated regions, i.e from $q_1$ to $q_4$.
% Although CMTs exist only in a statistical sense and single trajectories are not expected to follow average trajectories, this cycle is a characteristic feature of turbulence and exposes causality between the different configurations of the flow in an average sense
%\citep{ooi1999study}.
%The local topology of the velocity field can be conveniently studied through the second and third invariants of the velocity

To analyse the dynamics of the invariants, we decompose the CMTs
into the contribution of the different terms in the evolution equation of $Q$ and $R$, 
\begin{align}
\boldsymbol{\Phi}_E&=\{-3R,  2/3Q^2 \},\label{eq:DQ_1}\\
\boldsymbol{\Phi}_P&=\{\langle \mathcal{A}_{ij}\mathcal{H}_{ji}^P \rangle_{C}, \langle -\mathcal{A}_{ij}\mathcal{A}_{jk}\mathcal{H}_{ki}^P\rangle_{C}\},\label{eq:DQ_2}\\
\boldsymbol{\Phi}_M&=\{\langle \mathcal{A}_{ij}\mathcal{H}_{ji}^M \rangle_{C}, \langle -\mathcal{A}_{ij}\mathcal{A}_{jk}\mathcal{H}_{ki}^M \rangle_{C}\}\label{eq:DQ_3},
\end{align}
where $\boldsymbol{\Phi}_T=\boldsymbol{\Phi}_E+\boldsymbol{\Phi}_P+\boldsymbol{\Phi}_M$, 
and the normalisation with $t_Q$ has been dropped for simplicity.
This decomposition separates the strictly space-local dynamics of restricted Euler (RE) \citep[][]{cantwell1992exact}, 
$\boldsymbol{\Phi}_E$, which depends exclusively on $Q$ and $R$, and includes advection and the local action of the pressure Hessian,
from the non-local action of the pressure Hessian and the SGS model, which are included in $\boldsymbol{\Phi}_P$ and $\boldsymbol{\Phi}_M$ respectively.
This separation into local and non-local dynamics is convenient to justify the prevalence of direct over inverse cascades.
In order to compare the importance of the different terms, we consider the norm of the probability transport velocities, 
\begin{equation}
|\boldsymbol \Phi |=\sqrt{(t_Q^3\langle {D_tQ}\rangle_C)^2 + (t_Q^4\langle{D_tR}\rangle_C)^2}.
\label{equ:intensity}
\end{equation}
Statistics of $Q$ and $R$ and their CMTs have been compiled for two different times in the ensemble of realisations, $t_{stats}=(1 \pm 0.12)t_{inv}$, which correspond to the direct and inverse evolutions and are marked in figure \ref{fig:parameters_evolution}(a).
If we assume that the decay of the system is self-similar, conclusions drawn from this analysis should be independent of $t_{stats}$. The total probability transport velocity, $\boldsymbol{\Phi}_T$, has been calculated directly from the expression of the substantial derivative, computing separately the temporal and convective derivative.
%For the different contributions in (\ref{eq:DQ_1}, \ref{eq:DQ_2}, \ref{eq:DQ_3}) the explicit expression has been used for simplicity \citep[see][]{cantwell1992exact, meneveau2011lagrangian}. 
Appropriate numerical methods have been used in all the computations \citep{Lozanoduran2015}. Results show that the number of realisations in our database is sufficient to converge the statistics. 
%
%Descriptionf
%%

The different contributions to the  CMTs and the norm of the probability transport velocities in the inverse and direct evolutions are shown in figures \ref{fig:lyap_QR_con}(a-h) and  \ref{fig:lyap_QR_con}(i-l).
% In the direct evolution we observe a regular teardrop shape.
% Vieillefosse tail in $q_4$ and a higher probability of vortex stretching in $q_2$, which agrees well with the classical teardrop picture found in the literature.
%%
In the direct evolution, the CMTs develop an average rotating cycle characteristic of turbulent flows, which exposes causality between the different configurations of the flow in an average sense.
The CMTs rotate clockwise: from dominant vortex stretching to dominant vortex compression in the upper semiplane, to the Vieillefosse tail in $q_4$, and again to dominant vortex stretching in enstrophy-dominated regions.
 
The CMTs move from negative to positive $R$, and finally to the Vieillefosse tail, due to the effect of the RE dynamics. This trend would cause the appearance of infinite gradients in a finite time \citep{vieillefosse1984internal}, but the non-local effect of the pressure Hessian and the SGS model, or viscosity in the case of DNSs, prevent it by bringing the CMTs back to $q_3$ and restarting the cycle. For closed steady states, CMTs describe a closed cycle \citep{lozano2016multiscale},
while in our decaying flow they spiral inwards due to the action the SGS model, which contracts the probability distribution, resembling the action of the viscosity in DNSs \citep{meneveau2011lagrangian}.
The CMTs of the non-local component of the pressure Hessian evolve from positive to negative values of $R$, following the same behaviour observed by \citet{chevillard2008modeling} and \citet{meneveau2011lagrangian}.
The norm of the probability transport velocities reveals that the RE dynamics are mostly counteracted by the non-local component of the pressure Hessian in regions where $Q$ and $R$ are large. These observations are in agreement with \citet{luethi2009expanding}, and evidence a secondary role of the model in the dynamics of intense gradients.

In the inverse evolution we identify substantial changes. 
% In this analysis we must take into consideration the effect that the transformation $\boldsymbol{u}\rightarrow - \boldsymbol{u}$ has on the dynamics.
As explained in $\S$\ref{sec:num}, time-reversal changes only the sign of quantities which are odd with the velocity: $Q$ remains unaltered while $R$ changes sign, 
leading to an inverse teardrop shape.
The Vieillefosse tail lies now in $q_3$, forming an antitail,
and a higher probability of vortex compression appears in enstrophy-dominated regions.
This transformation also affects the CMTs,
and is most relevant in the effect of the SGS model, which now expands the probability distribution and leads to an average outward spiralling.
The behaviour of $\boldsymbol \Phi_E$ and $\boldsymbol \Phi_P$
is similar to the direct evolution in the upper semiplane,
but undergoes fundamental changes in the lower semiplane.
In the direct evolution, the non-local component of the pressure Hessian counteracts the RE dynamics, preventing the formation of intense gradients in the Vieillefosse tail.
Conversely, in the inverse evolution,  $\boldsymbol \Phi_P$ favours the growth of intense gradients in the antitail, while the RE opposes it.

This analysis compares the structure of the `attractor' with that of the `antiattractor', and can be simply derived by considering the effect that the transformation $\boldsymbol{u}\rightarrow - \boldsymbol{u}$ has on the dynamics of the velocity gradients. However,  we have identified in $\S$\ref{sec:per} inverse trajectories 
outside the antiattractor. We will show that, for these trajectories, this analysis yields non-trivial results.

% As noted before, the contribution of the non-local component of the pressure Hessian 
% is at least an order of magnitude higher in that region than the SGS model,
% suggesting that pressure forces play the most important role in sustaining the antitail.
% % In the inverse teardrop we find higher probability of vortex compression, which is known to cause the breakdown of vortices due to non-local pressure instabilities under certain conditions \citep{benjamin1962theory}.
% % In the case of the sustained inverse cascade, these compressed structures might persist in time due to the non-local action of the pressure forces \citep{verzicco1995steady}.
% We will later show that the pressure plays a crucial role in sustaining these structures in the inverse evolution.

\begin{figure}
  \begin{center}
  \vspace{18pt}
    \includegraphics[width=0.47\textwidth]{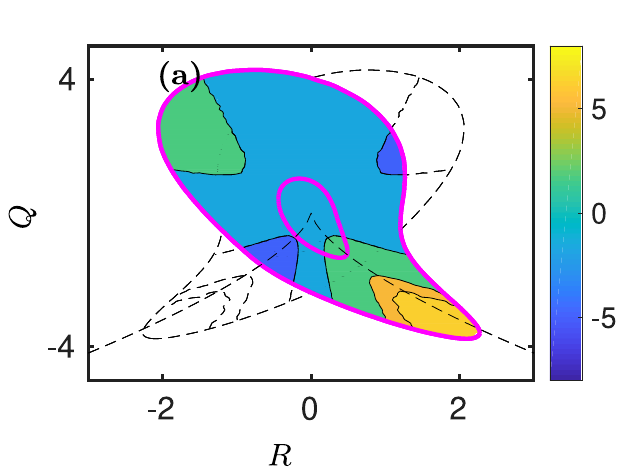}
    \includegraphics[width=0.445\textwidth]{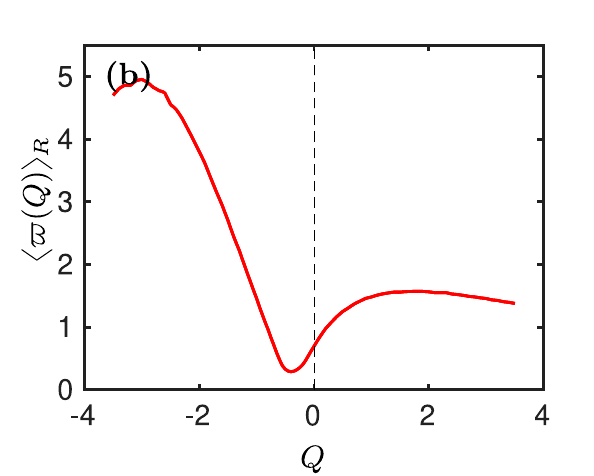}

      %\psfrag{X}{\hspace{-10pt}${\Delta}/\Delta_{g}$}
       \caption{(a) Colour plot of the asymmetry function $\varpi(Q,R)$. Isocontours contain $0.7$ and $0.85$ of the total data. Dashed lines represents the contours of the asymmetry function when $R\rightarrow-R$. (b) Asymmetry function averaged along the $R$ axis, $\langle\varpi(Q)\rangle_R$, for $R$ in the interval $(-2,2)$.
       Quantities are normalised using $t_Q={\langle Q^2 \rangle}^{1/4}$.
       } 
    \label{fig:entQR}
  \end{center}
\end{figure}

\begin{figure}
%   \begin{center}
\centering
  \vspace{7.5pt}
      \includegraphics[width=\textwidth]{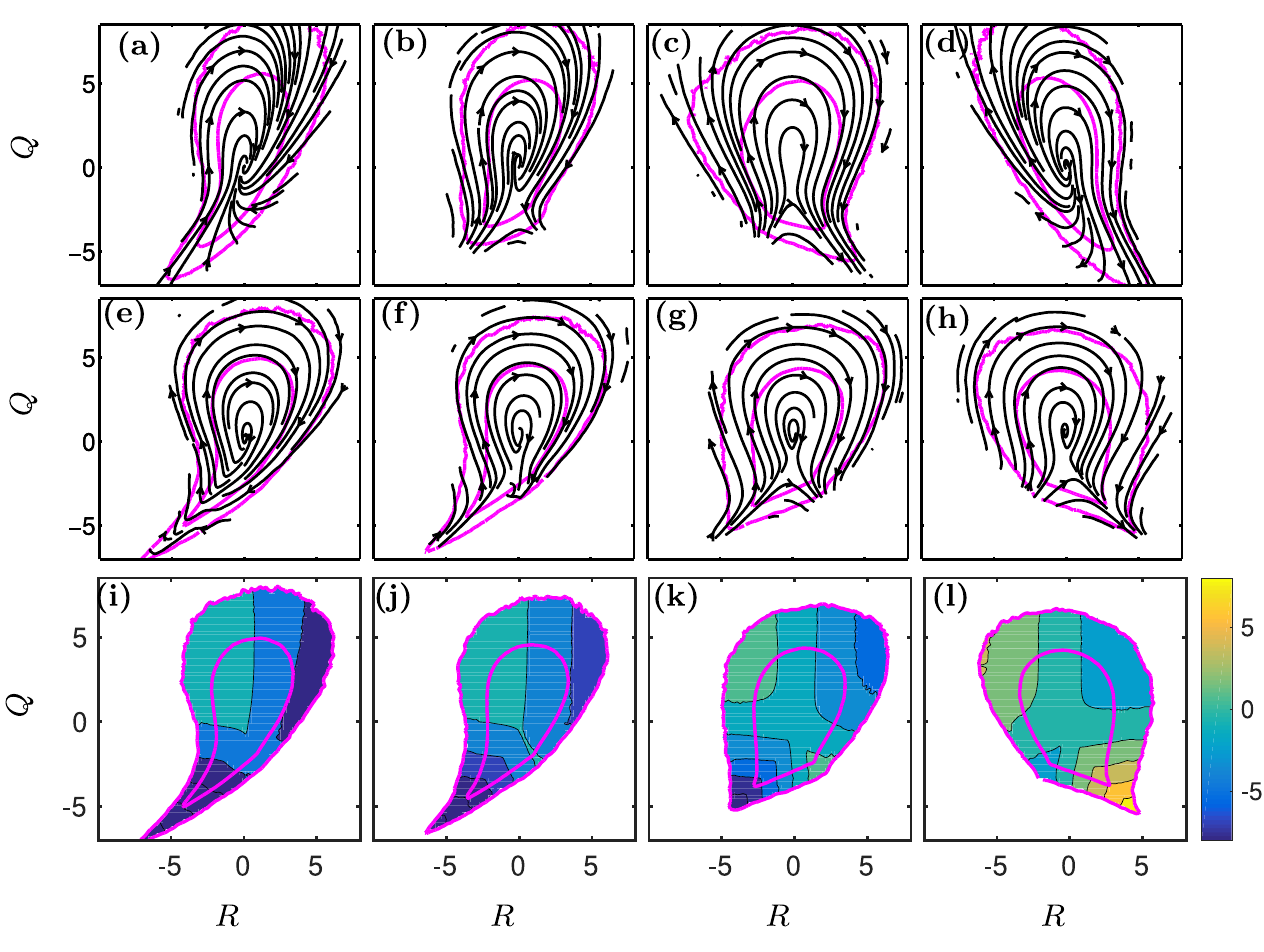}

          \caption{Conditional mean trajectories in the $Q$--$R$ plane of the inverse evolution without model (see $\S$\ref{sec:nomod}) for (a-d) the unfiltered case and (e-h) filtered at $\widecheck{\Delta}=5\Delta_g$, and (i-l) conditional average of $\Sigma$ at scale $\widecheck{\Delta}=5\Delta_g$, at different times:
          (a,e) $t/T_0=0$; (b,f) $t/T_0=0.1$; (c,g) $t/T_0=0.15$; (d,h) $t/T_0=0.25$. {\protect\color{magenta} \solid}, contours of the probability density function of $Q$ and $R$ containing $0.8$ and $0.9$ of the data. $Q$ and $R$ normalised with $t_Q=\langle Q(t_{inv})^2\rangle^{1/4}$.
          Conditional-averaged energy fluxes in (i-l) are normalised with the absolute value 
          of $\langle\Sigma\rangle$ at each time.
          }
    \label{fig:lyap_QR_nomod}
%   \end{center}
\end{figure}

\subsection{Asymmetry in the $Q$--$R$ space}
\label{sec:asy}
% As exposed in (\ref{eq:eoeo2}), the FR define entropy production by the ratio of probabilities of direct to inverse trajectories of duration $\Delta t$, where non-zero entropy production arises from the asymmetry in phase-space probabilities.
% In the spirit of the definition

% The differences in the $Q$-$R$ plane between the direct and inverse evolutions are significant,
% but are mostly identifiable in the lower semiplane. 
% where the asymmetry of the probability distribution is more pronounced.
The statistical irreversibility of the cascade is clearly manifest in the statistics of the invariants
through the asymmetry of the probability distribution with respect to $R=0$.
To characterize statistical irreversibility in the $Q$-$R$ plane, we use an asymmetry function 
%%%
\begin{equation}
\varpi(Q,R)=\log \frac{P^+(Q,R)}{P^-(Q,R)},
\label{eq:omega}
\end{equation}
%%%
where $P^+$ and $P^-$ denote the probability density of $Q$ and $R$ in the direct and inverse evolutions respectively. 
% Let us note that $\varpi(Q,R)$ is strictly zero for an inviscid truncated Euler flow in an equilibrium state.
Figure \ref{fig:entQR}(a) shows the distribution of $\varpi(Q,R)$, while \ref{fig:entQR}(b) 
shows the absolute values of the asymmetry function averaged along the $R$ axis,
\begin{equation}
\langle\varpi(Q)\rangle_R=\frac{\int|\varpi(Q,R)|\dd{R}}{\int \dd R}.
\end{equation}
Most of the temporal asymmetry of the velocity gradients is related to $Q<0$,
where the strain is dominant over the vorticity.
% The dynamics of enstrophy-dominated regions
% are statistically less affected by a time reflection
% than the dynamics of strain-dominated regions. 
In enstrophy-dominated regions, vortex stretching is on average dominant in the direct evolution,
but the probability of vortex compression is not negligible. 
Both processes also occur in the inverse cascade,
in which vortex compression is more probable but coexists with vortex stretching.
On the other hand, the structure of strain-dominated regions depends
strongly on the direction of the system in time.
Intense velocity gradients in $Q<0$ are either in $q_3$ or $q_4$, depending on the 
direction of the cascade, but not simultaneously in both quadrants.
The dynamics of strain-dominated regions 
are more fundamentally related to the direction of the system in time 
than the dynamics of enstrophy-dominated regions.

\subsection{Energy fluxes in the $Q$--$R$ plane: inverse evolutions outside the `antiattractor'}
\label{sec:spacelocal}

We now analyse the relation between the local structure of the flow and the energy cascade
by conditioning the statistics of the local energy fluxes to the invariants of the filtered velocity gradients. 
First, we study the experiments on the inverse cascade without model, which are presented in $\S$\ref{sec:nomod}.
These experiments provide non-trivial inverse evolutions outside the `antiattractor',
for which this analysis identifies the structures that support the inverse energy cascade.

We calculate the invariants of the filtered velocity gradients and their CMTs at scale $\widecheck{\Delta}=5\Delta_g$. 
We use a Gaussian filter (\ref{eq:gfil}), and calculate the CMTs using the substantial derivative of the filtered field, $\widecheck{D}_t=\partial_t+\widecheck{u}_i\partial_i$. We analyse these quantities in the inverse evolutions when the model is removed,
and connect them to the evolution of the local energy fluxes in physical space, $\Sigma(\boldsymbol x,t)$, and in scale space, $\Pi(k,t)$.
The analysis of $\Psi$ yields qualitatively similar results.   
%%%%%

Figures \ref{fig:lyap_QR_nomod}(a-h) show the probability distribution of the invariants of
the unfiltered (a-d), and filtered (e-h) 
velocity gradients, and the their CMTs, in the inverse evolution without model. 
Figures \ref{fig:lyap_QR_nomod}(i-l), show the averaged $\Sigma(5\Delta_g)$ conditioned to the invariants of the filtered velocity gradients.

Initially, the statistics of the filtered and unfiltered velocity gradients are similar, and resemble those of the inverse evolutions.
The average $\Sigma$ conditioned to $Q$ and $R$ is predominantly negative and most intense in $q_3$ and $q_1$, suggesting a relevant role of the antitail and of vortex compression in the inverse cascade.

At time $0.1T_0$ after the inversion, the antitail of the unfiltered
gradients has contracted, while the antitail of the filtered gradients remains unaltered.
We do not yet see changes in the upper semiplane, where the probability distribution remains similar to the initial state in both cases.
According to $\Pi(k,t)$ in figure \ref{fig:nomodel_1}(a), 
at this time the direct cascade has not yet regenerated at the small scales.
At the filter scale, $\Pi(k,t)$ and $\langle \Sigma \rangle$ are 
still negative, and  have a similar magnitude to the initial state. 
% The filter scale of this analys corresponds approximately in \ref{fig:nomodel_1}(a) to wavenumber $k\Delta_g=$

At $0.15T_0$, the unfiltered velocity gradients start developing dominant vortex stretching,
represented by a prominent lobe in $q_2$, and a regular Vieillefosse tail.
The antitail, although reduced, is still observable in the filtered gradients and $\langle \Sigma \rangle<0$. At this stage, as shown in \ref{fig:nomodel_1}(a), an inverse cascade at the filter scale, and a direct cascade in the small scales coexist. 
Finally, at $0.25T_0$, when the filtered velocity gradients develop structures in the Vieillefosse tail, the direct cascade regenerates at the filter scale and $\langle \Sigma \rangle>0$.

These results indicate a strong connection between the direction of the cascade and the orientation of the Vieillefosse tail. First, the conditional averages of $\Sigma$ in the $Q$-$R$ plane show that, at $0.25T_0$, when the energy cascade starts to regenerate at the filter scale, 
direct local energy fluxes are most intense in the $q_4$ quadrant. 
Second, the differences in the invariants of the filtered velocity gradients between $0.1T_0$ and $0.15T_0$ are only significant in the Vieillefosse tail, but the inverse energy fluxes are reduced by a half, $\langle\Sigma(0.15T_0)\rangle \simeq 0.5\langle \Sigma(0.1T_0)\rangle$.

The symmetry in the statistics of enstrophy-dominated regions reported in $\S$\ref{sec:asy} also holds during the transition from inverse to direct phase-space trajectories.  At the filter scale, the probability distribution or the CMTs in the upper semiplane does not substantially change during the regeneration of the energy cascade.

% nd, consequently, of dominant direct energy fluxes, is a consequence of the action of RE dynamics.
% In figure \ref{fig:lyap_QR_nomod_tau}(d) we observe that direct fluxes are intense in the Vieillefosse tail in its early development.
% The final shape of the probability distribution can be observed in figure \ref{fig:lyap_QR_tau_cond}(a), where we show averaged $\Sigma$ conditioned to $Q$ and $R$ for the direct cascade.  

\begin{figure}
%   \begin{center}
% \centering%
 \vspace{7.5pt}%
    \includegraphics[width=0.309\textwidth]{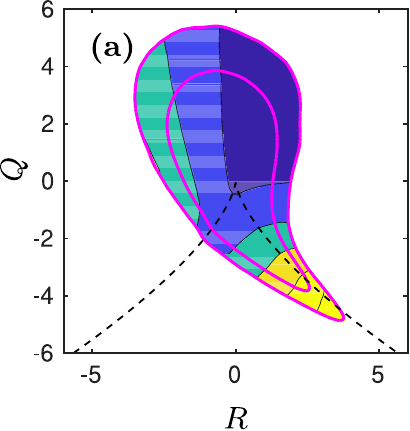}
    \includegraphics[width=0.3\textwidth]{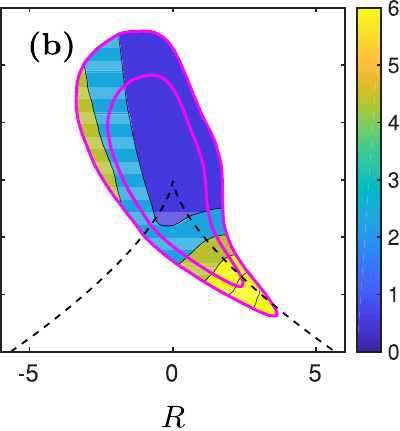}
    \includegraphics[width=0.348\textwidth]{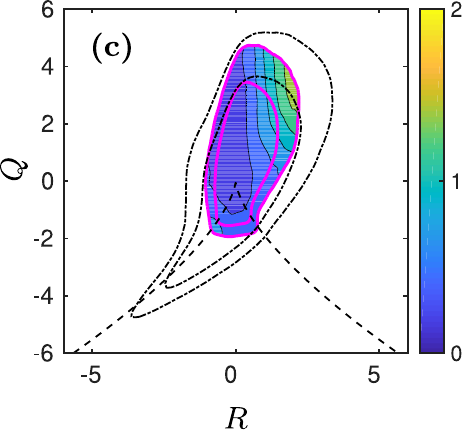}

% %   \centering%
% \vspace{-2pt}%%o
% \psfrag{Y}{\hspace{-13pt} $Q/{\langle Q_0^2 \rangle}^{1/2}$}%
%   \psfrag{X}{\hspace{-13pt}\raisebox{-4pt}{$R/{\langle Q_0^2 \rangle}^{3/4}$}}%
%    \includegraphics[width=0.95\textwidth]{QR_negative_all_taus.pdf}% 
%%%%%%%%
   \vspace{5pt}
          \caption{(a) Conditional average of $\Sigma$ at scale $\widecheck{\Delta}=5\Delta_g$ in the $Q$--$R$ plane of the filtered velocity gradients for the direct evolution with model.
          (b) Average $\Sigma$ over $Q$ and $R$ conditioned to $\Sigma>0$. 
          (c) Absolute value of the average $\Sigma$ over $Q$ and $R$ conditioned to $\Sigma<0$.
          Values of $\Sigma$ normalised with $\langle \Sigma \rangle$ at $t_{stats}$.
          {\protect\color{magenta} \solid}, contours of the probability density function of $Q$ and $R$ containing $0.85$ and $0.9$ of the data. In (c), \dotdashed marks isocontours of $Q$ and $R$ conditioned to negative $\Sigma$ for the inverse evolutions. The dashed line corresponds to $D=0$.
          $Q$ and $R$ normalised with $t_Q={\langle Q^2 \rangle}^{1/4}$.}
    \label{fig:lyap_QR_tau_cond}
%   \end{center}
\end{figure}

%Although

% We will show that intense inverse energy transfer events in the direct cascade are also related to an antitail, although this is precluded in the statistics due to the low probability of these events.
% To further characterize the relation between the Vieillefosse tail and the energy cascade,

\subsection{Energy fluxes in the $Q$--$R$ plane: direct evolutions in the turbulent `attractor'}

% ================================================================
\begin{table}
  \begin{center}
\def~{\hphantom{0}}
  \begin{tabular}{l}
    \hspace{10pt}\\
  \\[3pt]
  \\[3pt]
  $\langle \Sigma\rangle_{q_i}/\langle \Sigma \rangle$        \\[3pt]
  $\{ \Sigma \}_{q_i}/\langle \Sigma \rangle$  \\
  \end{tabular}
  \begin{tabular}{cccc}
   \multicolumn{4}{c}{All} \\
   \hline
    $q_1$ & $q_2$  &  $q_3$ & $q_4$  \\[3pt]
    $0.11$ & $\bf 1.05$  &  $0.81$ & $\bf 1.54$  \\[3pt]
    $0.01$ & $\bf 0.28$ & $0.17$  &  $\bf 0.54$ \\
  \end{tabular}
  \begin{tabular}{cccc}
  \multicolumn{4}{c}{`$+$'} \\
  \hline
    $q_1$ & $q_2$   &  $q_3$  & $q_4$  \\[3pt]
    $0.37$ & $\bf 0.89$ & $0.85$ & $\bf 1.3$  \\[3pt]
    $0.03$ & $\bf 0.27$ & $0.16$ & $\bf 0.52$ \\
  \end{tabular}
  \begin{tabular}{cccc}
  \multicolumn{4}{c}{`$-$'} \\
  \hline
    $q_1$ & $q_2$  &  $q_3$ & $q_4$  \\[3pt]
    $\bf 1.16$ & $0.72$ & $0.91$ & $0.89$  \\[3pt]
    $\bf 0.56$ & $0.11$ & $0.17$ & $0.16$ \\
  \end{tabular}
  \caption{Conditional averages of energy transfer and contributions to 
  the total energy fluxes of each quadrant in the $Q$--$R$ plane during the direct evolution. 
  In the first table we have considered all data,
  in the second only direct energy transfer events,
  and in the last only backscatter events.
  Quantities are normalised with the average $\Sigma$ in the first case,
  and with the averages conditioned to positive and negative fluxes in the other two cases.}
  \label{tab:quadrants}
  \end{center}
\end{table}
% ==========================================================

We perform the same analysis for the direct evolutions with SGS model.
Figure \ref{fig:lyap_QR_tau_cond}(a) shows the average $\Sigma$ conditioned to $Q$ and $R$.
To discriminate between direct and inverse energy transfer,
we condition the probability distribution of $Q$ and $R$ to the sign of
the local energy fluxes, and present the results in figures \ref{fig:lyap_QR_tau_cond}(b-c).
% characterize the importance of each quadrant in the energy cascade,
We quantify in table \ref{tab:quadrants} the relevance of each quadrant to the energy cascade 
by calculating the average of $\Sigma$ conditioned to $Q$ and $R$ belonging to a quadrant,
$\langle \Sigma \rangle_{q_i}=\langle \Sigma |(Q,R)\in q_i\rangle$,
and the total contribution of each quadrant to the energy fluxes, $\{\Sigma\}_{q_i}=\langle \Sigma \rangle_{q_i}P_{q_i}$, where $P_{q_i}$ is the probability that $Q$ and $R$ belong to $q_i$, so that $\langle \Sigma \rangle=\sum_{i=1,4}\{\Sigma\}_{q_i}$.
We have also calculated these quantities conditioning the averages to the direction of 
the energy fluxes, being
\begin{equation}
\langle \Sigma \rangle^+_{q_i}=\langle \Sigma |\Sigma>0,(Q,R)\in q_i\rangle,
\end{equation}
the average energy transfer conditioned to the quadrant $q_i$ and to $\Sigma>0$.
Accordingly $\{\Sigma\}^+_{q_i}=\langle \Sigma \rangle^+_{q_i}P^+_{q_i}$ is the total contribution
of the quadrant $q_i$ to the positive energy fluxes.
Here $P^+_{q_i}$ is the probability that $Q$ and $R$ belongs to $q_i$ and that $\Sigma>0$.
Statistics conditioned to negative fluxes are marked with a minus superscript.
% The value of these quantities are gathered in table \ref{tab:quadrants}.

The most intense direct energy transfer events are related to structures in the Vieillefosse tail, and $q_4$ has the highest average energy transfer, $\langle \Sigma \rangle_{q_4}\sim1.5\langle \Sigma \rangle$, and contributes with approximately $50\%$ of the total energy transfer. Also structures in $q_2$ are relevant to the cascade, with a contribution of around $30\%$ to the total fluxes. Conversely, the contribution of $q_1$ to direct energy transfer is negligible. 
When we condition these statistics to direct energy fluxes we obtain qualitatively similar results.
The statistics of $Q$ and $R$ conditioned to positive $\Sigma$ yield almost a complete teardrop, except for a lower probability of events in $q_1$.

When we condition to negative energy transfer events, the average inverse energy transfer is roughly similar in the four quadrants,
but the $q_1$ quadrant contributes the most to the total, with a $56\%$.
This is a consequence of inverse energy transfer events being mostly located in $q_1$.
As shown in figure \ref{fig:lyap_QR_tau_cond}(c), the statistics of $Q$--$R$ conditioned to inverse energy transfer resemble an inverse teardrop without the antitail.
In the same figure we have plotted the probability density function of $Q$ and $R$ conditioned to inverse energy transfer in the inverse evolutions.
By time symmetry, the antitail in $q_3$ contributes the most to inverse energy transfer in the inverse cascade, but this tail is absent in the direct evolution. 

\subsection{An entropic argument for the prevalence of direct energy fluxes}
\label{sec:sub_sec}

% We propose an entropic argument to explain the low probability of inverse cascades.
We have presented in the previous section compelling evidence that topologies in the antitail are strongly connected to
the inverse cascade, but that these topologies have negligible probability in the direct cascade.
Here we provide arguments that explain the low probability of these topologies, 
and, consequently, of inverse cascades.

Consider the evolution of $R$ in regions where the strain is dominant over the enstrophy, 
$|S|^2\sim\langle|S|^2\rangle\gg|\omega|^2$. Neglecting the contribution of the enstrophy, we obtain
\begin{equation}
R\approx-\tfrac{1}{3}S_{ij}S_{jk}S_{ki}=-\alpha_1\alpha_2\alpha_3,
\end{equation}
where $\alpha_1\leq\alpha_2\leq\alpha_3$ are the eigenvalues of the rate-of-strain tensor. 
This is a good approximation to the Vieillefosse tail and the antitail, 
where $|S|^2\sim100|\omega|^2$.
Due to compressibility $\alpha_1 + \alpha_2 + \alpha_3=0$, and the sign of $R$ is determined by the sign of the intermediate eigenvalue.
Topologies in the Vieillefosse tail are characterised by $\alpha_2>0$,
while in the antitail $\alpha_2<0$.
We describe the evolution of $R$ in terms of the evolution of $\alpha_2$, whose equation reads
\begin{equation}
{D_t\alpha_2}=-\alpha_2^2-(\frac{\Lambda}{3}+\Gamma_2),
% {D_t\alpha_2}&=-\alpha_2^2-(\frac{\Delta}{3}+\Gamma_2), \\
% {D_t\alpha_3}&=-\alpha_3^2-(\frac{\Delta}{3}+\Gamma_3), 
\label{eq:RE_strain_end}
\end{equation}
where $\Lambda=\partial_{kk}p\approx-\tfrac{1}{2}|S|^2=-\alpha^2_1-\alpha^2_2-\alpha^2_3$ is the trace of the pressure Hessian, which is local, and $\Gamma_2$ is the contribution of $\mathcal H^P_{ij}=\partial_{ij} p-\tfrac{1}{3}\partial_{kk}p\delta_{ij}$ to the evolution of $\alpha_2$, which is non-local. We have discarded quadratic terms in the vorticity and the action of the SGS model, which we show above to have a negligible contribution to the dynamics of intense gradients \citep{nomura1998structure}.
Disregarding the action of the non-local component of the pressure Hessian, $\Gamma_2$, the equation reads
\begin{equation}
D_t\alpha_2=\frac{1}{3}(\alpha_1^2+\alpha_3^2-2\alpha_2^2)>0,
\label{eq:RE_strain_11}
\end{equation}
where we have used that, from their definition,  $\alpha_1^2>\alpha_2^2$ and $\alpha_3^2>\alpha_2^2$. 
The evolution of any initial value of $\alpha_2$ results in the growth of the intermediate eigenvalue, leading to positive values of $R$.
This is the effect of the RE dynamics, which depletes the antitail in the direct cascade.
Left to itself, it generates intense strain in the Vieillefosse tail, and is independent on the non-local structure of the velocity gradients.
From (\ref{eq:RE_strain_11}) we conclude that the only inertial mechanism capable of preventing
the rate-of-strain tensor from developing a positive intermediate eigenvalue 
is the non-local action of the pressure Hessian through $\Gamma_2$.
This analysis is consistent with the results presented in $\S$\ref{sec:dyninv}, 
where we show that this term sustains the antitail in the inverse cascade.

The non-local pressure Hessian depends on the complete flow field. 
In the absence of boundaries it can be expressed as a singular integral, which depends on all the points of the domain \citep{speziale1991modelling,ohkitani1995nonlocal}.
As a consequence, the non-local pressure Hessian is in general decorrelated from the local dynamics of the velocity gradients \citep{she1991structure}.
%
% In figure \ref{fig:hess}, we show the joint probability distribution of the 
% of the two components of $\boldsymbol \Phi_P=\{\mathcal{A}_{ij}\mathcal{H}^P_{ji},-\mathcal{A}_{ij}\mathcal{A}_{jk}\mathcal{H}^P_{ki}\}$ in the center of the the $Q$--$R$ plane, and in the Vieillefosse tail during the inverse and direct evolution.
% In regions where gradients are weak the non-local action of the pressure Hessian appears decorrelated from the local dynamics of the velocity gradients.
% In the Vieillefosse tail this correlation increases.
%
%
\begin{figure}
\centering
    \includegraphics[width=0.52\textwidth]{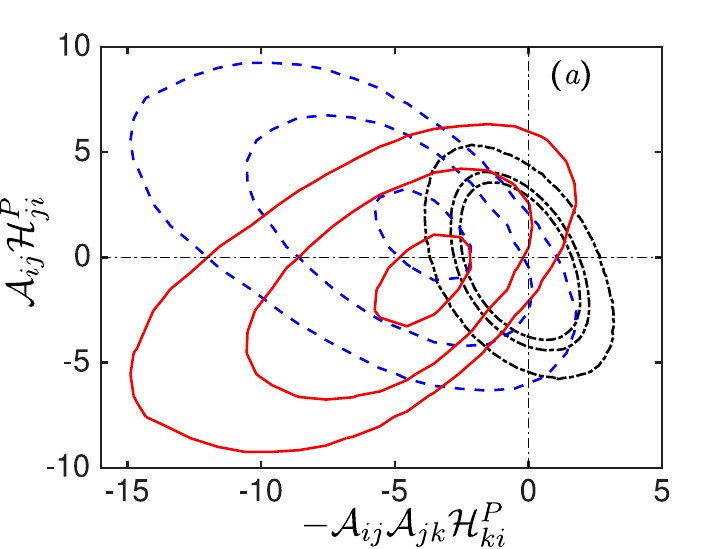}
    \includegraphics[width=0.38\textwidth]{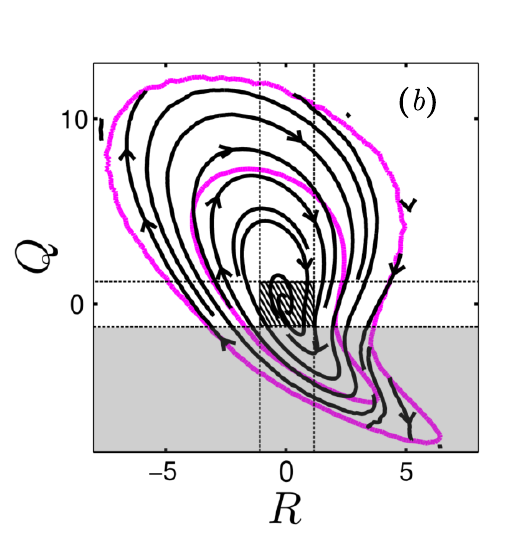}

          \caption{ (a) Joint probability density function of the components of the probability transport velocities due to the non-local part of the pressure Hessian, $\{\mathcal{A}_{ij}\mathcal{H}^P_{ji},-\mathcal{A}_{ij}\mathcal{A}_{jk}\mathcal{H}^P_{ki}\}$, 
          conditioned to different areas in the $Q$-$R$ plane:
          \dotdashed, $|Q|<1$ and $|R|<1$, dashed region in (b);
          {\color{blue}\dashed}, $Q<-1$ in the direct evolution, shaded region in (b);
          {\color{red}\solid}, $Q<-1$ in the inverse evolution, shaded region in (b). 
          Quantities normalised with $t_Q$.
          }
          \label{fig:hess}
\end{figure}
%
% A similar argument can be used to explain 
% 
% The non-local component of the pressure Hessian is decorrelated 
% from the local dynamics of the velocity gradients in regions where the strain is weak \citep{she1991structure}.
% We study the 
Figure \ref{fig:hess}(a) shows the joint probability distribution
of the two components of $\boldsymbol \Phi_P=\{\mathcal{A}_{ij}\mathcal{H}^P_{ji},-\mathcal{A}_{ij}\mathcal{A}_{jk}\mathcal{H}^P_{ki}\}$ in the centre of the the $Q$--$R$ plane, and in the Vieillefosse tail during the inverse and direct evolutions. See figure \ref{fig:hess}(b) for reference.
% First in the center of the $Q$-$R$ plane,where $|Q|<\langle Q^2\rangle^{1/2}$ and $|R|<\langle Q^2\rangle^{3/4}$, and second, in the Vieillefosse tail, $Q<-\langle Q^2 \rangle^{1/2}$. 
In regions where gradients are not strong, the non-local component of the pressure Hessian appears decorrelated from the local dynamics, with a large scatter of the data with respect to the mean.
In the Vieillefosse tail, the non-local component of the pressure Hessian 
counteracts the RE dynamics and the action of the non-local component of the pressure Hessian 
over $Q$ and $R$ appears more correlated. When time is reversed, these correlations are necessary to sustain the anti-tail. As shown in \ref{sec:spacelocal}, the contraction of the antitail is the 
first identifiable process in the transition from the inverse to the direct cascade, 
which indicates that these correlations are quickly destroyed under perturbations.

While the generation of topologies in the Vieillefosse tail is a direct consequence
of the local self-amplification of the rate-of-strain tensor, 
the generation of topologies in the antitail requires 
a global configuration of the complete flow field 
that counteracts this amplification
through the non-local action of the pressure Hessian.
This scenario appears intuitively unlikely, suggesting 
that only specially organised flows are able to produce such action \citep{popper1956arrow}.
This is the case of the initial conditions used for the inverse evolutions, 
which conserve all the information of the direct cascade process.
In particular, they retain the correlations between the rate-of-strain tensor and the non-local contribution of the pressure Hessian in the Vieillefosse tail, 
thus inhibiting the breakdown of the antitail in the inverse cascade.

\section{Conclusions}
\label{sec:conc}

We have studied a microscopically reversible turbulent system constructed using 
a reversible sub-grid scale (SGS) model to explore the energy cascade as an entropy-driven process. 
In this system inverse and direct energy cascades over the full domain are possible, 
but only the latter are experimentally observed due to the statistical irreversibility
of the dynamics in the inertial range.
In this paper, we explain the origin of this irreversibility by focusing on the 
dynamics of the energy cascade in physical space. 
% connecting information-entropy production with turbulence mechanisms.

% In $\S$\ref{sec:exp} we have conducted different \refb{numerical experiments} on reversible turbulence.
We show in $\S$\ref{sec:nomod} that, in this reversible system,
sustained inverse cascades are a consequence of inertial mechanisms,
while the SGS model only acts as a source of energy. 
By reversing the energy cascade of a system with an irreversible SGS model,
we \refc{present strong evidences that} %proven 
microscopic reversibility is a fundamental property of inertial dynamics,
independent of the dissipation.

In $\S$\ref{sec:per}, we characterise the distribution of inverse trajectories in phase space.
By perturbing inverse phase-space trajectories within the `antiattractor', i.e the set of turbulent flows generated by evolving direct cascades backwards in time,
we show that inverse evolutions exist in a wide region of phase-space, separated from the direct evolutions. The perturbed inverse trajectories diverge from the original trajectories, and eventually evolve towards the turbulence `attractor', which is composed almost exclusively by trajectories with a direct energy cascade. 
Inverse cascades are `unstable'. They can only be observed for a finite time, and it is extremely unlikely that 
they develop spontaneously.

% \refc{Moreover, the fluctuations of the average sub-grid dissipation in the turbulent attractor are much smaller than the ensemble average, and therefore 
% far from negative values. 
% Since this quantity reflects the fluctuations of the energy fluxes at the grid filter, 
% and must be negative for an inverse cascade to survive, 
% this result supports the extremely low probability that the
% system spontaneously escapes the attractor and develops a sustained inverse cascade.}

% Moreover, the fluctuations of the volume-averaged sub-grid dissipation in the turbulent attractor are much smaller than the ensemble average, indicating that there is a extremely low probability that the
% system spontaneously escapes the attractor and develops an average inverse cascade.

% Although average inverse cascades are highly unlikely, local inverse energy transfer
% is a well known feature of turbulent flows. 
In order to connect phase-space dynamics with the energy cascade, 
we have studied the spatial structure of the local energy fluxes in physical space using two common 
definitions of the fluxes, which are calculated at different scales by filtering the velocity field.
The auto-correlation length of these fluxes is of the order of the filter width, and, when volume-averaged for different sizes, their statistics are 
consistent with a space-local process.
Although the two fluxes have similar mean, they are very different
pointwise, particularly in the intensity and probability of backscatter.
These differences disappear when the fluxes are averaged over volumes of the order of the filter width cubed.
We conclude that the energy cascade is approximately local in physical space and robust to the definition of the energy fluxes on an average sense.

We have extended this analysis by studying the spatial correlation between local energy fluxes at different scales.
The correlation between fluxes at scales separated by a factor of $2$ is substantial, 
but significantly decreases when scales are separated by a factor of $4$,
corroborating the reported scale locality of the cascade \citep{eyink2005locality,eyink2009localness,domaradzki2009locality,cardesa2017turbulent}.
When conditioning these correlations to the sign of the fluxes, 
we find that they are strong for direct energy fluxes but weak for energy backscatter,
revealing that, unlike direct energy transfer, backscatter is very shallow in scale space.
Moreover, backscatter is very sensitive to the choice of the fluxes, 
and disappears when volume-average over volumes of the order of the cubed filter width,
suggesting that it represents spatial fluxes at the scale of the filter rather than 
an inverse energy cascade involving interactions at different scales.

% These evidences suggest that the dynamical significance of backscatter to the energy cascade is limited.
% Although backscatter events are present in the statistics of the local energy fluxes, 
% they do not seem to imply the presence of an inverse cascade.
% The differences between the two definition of the fluxes, which are particularly evident in the frequency and intensity of
% backscatter events, disappear after volume-averaging, partially because of the reduction of backscatter events.
% \refc{Let us note that the difference between the two definitions is the divergence of a spatial flux,
% which produces no net energy transfer and, as we have shown, cancels out when volume-averaged over rather small neighbourhoods.
% This spatial flux is scale-local and free from large-scale effects such as sweeping.
% Altogether, the strong scale-locality of backscatter, the depletion of backscatter events when volume-averaged,
% and the sensitivity of backscatter statistics to the choice of the fluxes raises some doubts as to whether this phenomenon 
% is as relevant to the turbulent energy cascade as previously thought.
% We suggest that it could be an artifact of the particular definition of the energy fluxes, 
% and therefore of what we choose to consider scale-local and scale non-local interactions.}
Although the present work strongly emphasises the bidirectionality of the cascade,
our results \refc{suggest} that the dynamical significance of backscatter to the energy cascade is limited.
Following the local fluctuation relations \citep{michel2013local}, 
we have attempted to estimate the probability of inverse energy cascades
over the complete domain by quantifying the probability of volume-averaged inverse energy transfer events,
but our results are not conclusive because these events disappear almost completely for small averaging volumes. 
Turbulence is sufficiently far from equilibrium as to preclude 
the observation of local inverse cascades, even in reduced regions of space.
Although some representations of the energy fluxes can be negative locally, 
this does not seem to imply a local inverse energy cascade that develops against the 
tendency of the system towards equilibrium.   

% Evidences of the limited relevance of backscatter to the cascade are provided by the good agreement between LESs and DNSs, 
% despite the inability of available sub-grid models to reproduce this phenomenon.
% \refc{These considerations} are restricted to the energy cascade in isotropic turbulence,
% and may not apply in different conditions.
% For instance, in wall-bounded flows, some energy containing-scales are larger than the energy-producing
% scales, and backscatter is probably associated to an inverse cascade of energy that feeds the former, 
% due possibly to the presence of strong anisotropies in these flows \citep{piomelli1991subgrid}.

Finally, we have compared the structure of the direct and inverse cascades through the study of the invariants of the velocity gradient tensor, $Q$ and $R$, and its conditional mean trajectories (CMTs). 
In the direct evolutions, the joint probability density function of $Q$ and $R$ forms the classical teardrop, with dominant vortex stretching in $Q>0$ and a \citet{vieillefosse1984internal} tail  
in $Q<0$, while in the inverse evolutions $Q$ and $R$ form an inverse teardrop, with dominant vortex compression in $Q>0$ and an antitail in $Q<0$.  
We have quantified these differences using an asymmetry function in the $Q$--$R$ plane,
which reveals that most of the temporal asymmetry is located in the $Q<0$ semiplane,
suggesting a strong connection between the structure of the rate-of-strain tensor and the direction of the system in time. 
In the direct cascade the rate-of-strain tensor has predominantly one negative and two positive eigenvalues,
while in the inverse cascade it has two negative and one positive eigenvalues.
%and the topology of intense strain events is strongly conditioned to the direction of the cascade.
On the other hand, it is more challenging to distinguish the 
inverse from the direct cascade by only considering the dynamics of intense enstrophy regions:  
\refb{the stretching and compression of intense vorticity are present in both the direct and the inverse cascade.}
% vortex stretching and compression are common to both the direct and the inverse cascade.

The CMTs convey a similar picture. They display similar behaviour in $Q>0$ regardless of the direction of the cascade,
but undergo fundamental changes in $Q<0$ when the system is reversed in time.
In the direct evolution, the non-local action of the pressure Hessian opposes the formation of 
intense gradients in the Vieillefosse tail, counteracting the restricted Euler (RE) dynamics, 
while in the inverse cascade it favours the formation of the antitail, also counteracting the RE dynamics.

We have extended this analysis to inverse phase-space trajectories outside the antiattractor.
The transition from an inverse to a direct cascade is marked by the disappearance of the anti-tail and
the appearance of a regular Vieillefosse tail, but the statistics of enstrophy-dominated regions do not change substantially in this transition.

To further corroborate the strong connection between the Vieillefosse tail and the direction of the cascade, we have conditioned the statistics of local energy fluxes to the invariants of the filtered velocity gradient tensor at the same scale.
In the direct evolutions, the most intense energy transfer events are related to structures in the Vieillefosse tail, which contribute the most to the total energy fluxes. On the other hand, the antitail is responsible for most inverse energy transfer in the inverse evolutions.

% In the turbulent attractor, the intensity of backscatter does not depend strongly on the underlying topology, but most energy backscatter is related to dominant vortex compression. 
% These events are much weaker than the inverse energy transfer events in the antiattractor,
% which are related to structures in the antitail. 

In view of these results, we point to the negligible probability of topologies in the antitail as the cause of the low probability of inverse cascades,
and propose an entropic argument to explain this fact.
The generation of topologies along the Vieillefosse tail is encoded in the point-wise interaction of the velocity gradients, and is strictly local. 
Any possible organisation of the vorticity vector and the rate-of-strain tensor will generate topologies in the Vieillefosse tail under 
the action of RE dynamics.
On the other hand, the formation and maintenance of the antitail requires the global action of the non-local component of the pressure Hessian, 
which involves a large number of degrees of freedom, to generate intense strain with a negative intermediate eigenvalue.
% Due to non-locality, the dynamics of the non-local component of the pressure Hessian depend on a large number of degrees of freedom.
While the dynamics of any flow field, even random ones, leads to the generation of a Vieillefosse tail due to RE dynamics, only very special flow fields are sufficiently organised so as to counteract the RE dynamics and form an antitail.
\refb{Assuming that global fluctuation relations \citep{evans2002fluctuation} apply to turbulent flows,
the probability of observing such organized states is inversely proportional to the exponential of the number of degrees of freedom of the system,
which explains why stochastic and deterministic reduced-order models of the cascade exhibit statistical irreversibility
despite their reduced number of degrees of freedom \citep{biferale2003shell,meneveau2011lagrangian}.
% and  the probability of inverse cascades is negligible even for restricted regions of the flow. 
}

% Let us note that this argument applies also to the small-scale dynamics of turbulence.
% \citet{betchov1956inequality} proposed that the positive sign of the average strain self-amplificaion and vortex stretching, $-\langle S_{ij}S_{jk}S_{ki}\rangle=3/4\langle \omega_{i}S_{ij}\omega_j\rangle>0$, are a consequence of the balance between these terms and the dissipation of the velocity gradients.
% However, as stated by \citet{tsinober1998concentrated}, this argument is questionable since it puts the consequences before the causes. 
% Our reasoning suggests that the self-amplification of the velocity gradients, and therefore the generation of dissipation, is also entropic in nature.

Our results stress the importance of the rate-of-strain tensor on the dynamics of the energy cascade, 
\refb{ and agree with other studies that suggest that the role of vorticity in turbulence dynamics is perhaps overemphasized 
\citep{tsinober1998concentrated,carbone2020vortex,johnson2020energy}.}
While a change in the sign of vorticity does not change the fundamental dynamics of vortices, for instance in the \citet{burgers1948mathematical} model,
a change in the sign of the rate-of-strain tensor reverses the evolution of both the enstrophy and the strain in the inviscid case. 
This is evident in the evolution equations of the strain and the enstrophy, (\ref{equ:ss}-\ref{equ:ome}), which are odd in the rate-of-strain tensor and even in the vorticity vector.
Although the rate-of-strain tensor and the vorticity vector are not independent, but tied by kinematic relations, 
and changing the sign of one changes the sign of the other, it is the rate-of-strain tensor that reflects most of the statistical irreversibility of turbulent flows.
A local observer wishing to establish the direction of the cascade in the inertial scales of turbulence 
should focus on the dynamics of the rate-of-strain tensor, rather than on the dynamics of the vorticity vector.

% % ...................................................................
% \jj{{\em (Coming from \S2, adapt)} The second question is a more fundamental one. Since the
% inverse cascade represents phase-space trajectories with characteristics different from the
% observable turbulence, it is natural to consider the existence of other possible phase-space
% trajectories with direct energy cascades different from the turbulent one. The reproducible
% structure of turbulent flows suggests that not all trajectories towards equilibrium are
% equally probable, but that the system approaches equilibrium following a special set of
% trajectories with a well-defined, recurrent organised structure. The second question
% addresses how the system approaches equilibrium: what determines the selection of turbulent
% trajectories among all possible trajectories towards equilibrium? Although this second
% question is beyond the scope of this paper, we present a description of the characteristic
% structure of the energy cascade, and its relation with energy fluxes and entropy production,
% which might help determine the principles that rule the selection of turbulent evolutions. }
% % ...................................................................

\section*{}
This work  was supported by the European Research Council COTURB project ERC-2014.AdG-669505.   

\section*{Declaration of interests}
The authors report no conflict of interest.

%\appendix

%\bibliography{jfm-instructions}
\bibliographystyle{jfm}

\bibliography{draft}


\end{document}